\documentclass{jaa}
\usepackage{natbib}
\usepackage{graphicx}
\usepackage[english]{babel}

\usepackage{amsmath}

\usepackage[colorlinks=true, allcolors=blue]{hyperref}
\usepackage{subcaption}

\usepackage{soul}

\begin{document}\sloppy

\title{A summary of instruments proposed for observing pulsating variables from the Mt. Abu Observatory}


\author{Anwesh Kumar Mishra\textsuperscript{1,2,*}, Deekshya Roy Sarkar\textsuperscript{1}, Prachi  Prajapati\textsuperscript{1}, Alka Singh\textsuperscript{1}, Prashanth K. Kasarla\textsuperscript{1} and Shashikiran Ganesh\textsuperscript{1}}
\affilOne{\textsuperscript{1}Physical Research Laboratory, Ahmedabad, India.\\}
\affilTwo{\textsuperscript{2}Indian Institute of Astrophysics, Bangalore, 560034, India.}


\twocolumn[{

\maketitle

\corres{mishraanwesh@gmail.com}

\msinfo{1 January 2015}{1 January 2015}

\begin{abstract}
Pulsating variables play a significant role in shaping modern astronomy. Presently it is an exciting era in observational study of variable stars owing to surveys like OGLE and TESS. The vast number of sources being discovered by these surveys is also creating opportunities for 1-2 meter class telescopes to provide followup observations to characterise these. We present some initial observations of type-II cepheids from the Mt. Abu observatory and highlight the need for dedicated observing runs of pulsating variables. We also present optical designs for a number of suggested instruments for the Mt. Abu observatory that will contribute towards this goal. We present designs that are fairly simple and yet take due benefit of the unique telescopes and facilities present at the observatory.
\end{abstract}


}]


\doinum{12.3456/s78910-011-012-3}
\artcitid{\#\#\#\#}
\volnum{000}
\year{0000}
\pgrange{1--}
\setcounter{page}{1}
\lp{1}

\section{Introduction: pulsating variables}
Pulsating variables are typically stars that show fluctuations in their intrinsic brightness in a periodic manner. These stars are typically found in specific regions of the H-R diagram known as instability strip. Pulsating variables are typically stars in late stages of stellar evolution that have evolved off the main sequence and are passing through the instability strip. These stars include short period variables such as RR Lyrae, to classical cepheids and type-II cepheids (1-20 days) and then very long period variables such as RV Tau and Mira type variables (\cite{catelan2015pulsating}). Pulsating variables have been at the heart of a few significant debates in astrophysics. One such debate is concerning the Cepheid mass problem(\cite{stobie1969cepheid}, \cite{rodgers1970masses}, \cite{fricke1972masses} ), which focuses on the gap between mass estimates of cepheids from evolutionary and pulsational models. Attempts to resolve this problem have involved improved models of $H^-$ opacity (\cite{rogers1992radiative}, \cite{seaton1994opacities}) and invoking additional pulsation driven mass loss mechanisms for these stars (\cite{neilson2008enhancement}). At present, this problem still persists (albeit at a smaller proportion) and needs both theoretical and observational progress (\cite{weiss2012stellar}). Pulsating variables also play a crucial role in determining the Hubble constant and are at the centre of recent studies on Hubble tension (\cite{freedman2021measurements}, \cite{mortsell2022hubble}, \cite{woods2023jwst}). Apart from this, the pulsational mechanisms of these stars have also been an interesting and active area of study (\cite{buchler2009stat}).

The present era in variable star observation is that of large-scale discovery. Surveys such as OGLE\footnote{https://ogle.astrouw.edu.pl/}(\cite{udalski2015ogle}) and ASASSN\footnote{https://www.astronomy.ohio-state.edu/asassn/}(\cite{shappee2014all}) have revealed a large number of such sources and are continuously discovering more. This is in addition to an already significant number of sources catalogued by the General Catalogue of Variable Stars(GCVS, \cite{samus2017general}). At the same time, new mode of observations are emerging. With progress in  asteroseismology(\cite{aerts2021probing}, \cite{joshi2015asteroseismology}) it is now possible to probe the interior of stars. Precise period measurements such as by \cite{csornyei2022study} have made it possible to detect small changes of pulsational period as a measure of rate of evolution. Surveys such as TESS\footnote{https://tess.mit.edu/} (\cite{ricker2015transiting}) are likely to contribute significantly towards these type of observing modes of variables. Considering the large number of sources already catalogued and new and interesting sources likely to be discovered it is necessary for optimised observing programs for follow up observations on these sources.

One important aspect of followup observations for these sources is by infrared photometry. Typically, most pulsating variables are red and evolved stars and are therefore brighter in the infrared than in the visible. However, observation in the infrared wavelength range is found to be lacking for these type of sources. This is partly due to sparsity of infrared instruments (particularly for Indian astronomers) and also partly due to persistence issues in HgCdTe arrays that prevent them from pointing at bright sources.  It is strongly desirable to observe variable stars such as delta cepheids, type-II cepheids and long period variables in infrared( 900-2400 nm, sometimes known as SWIR). The period luminosity relation in infrared is "tighter" i.e. generally there is less scatter in the observed P-L laws in infrared wavelengths compared to visible (\cite{persson2004new}). Further, by monitoring the phase lag between the optical and infrared lightcurves, it may be possible to probe the pulsation mechanism of variable stars (\cite{bhardwaj2017comparative}, \cite{bhardwaj2022rr}).

\begin{figure}
    \centering
           
       
        \includegraphics[width=0.49\textwidth]{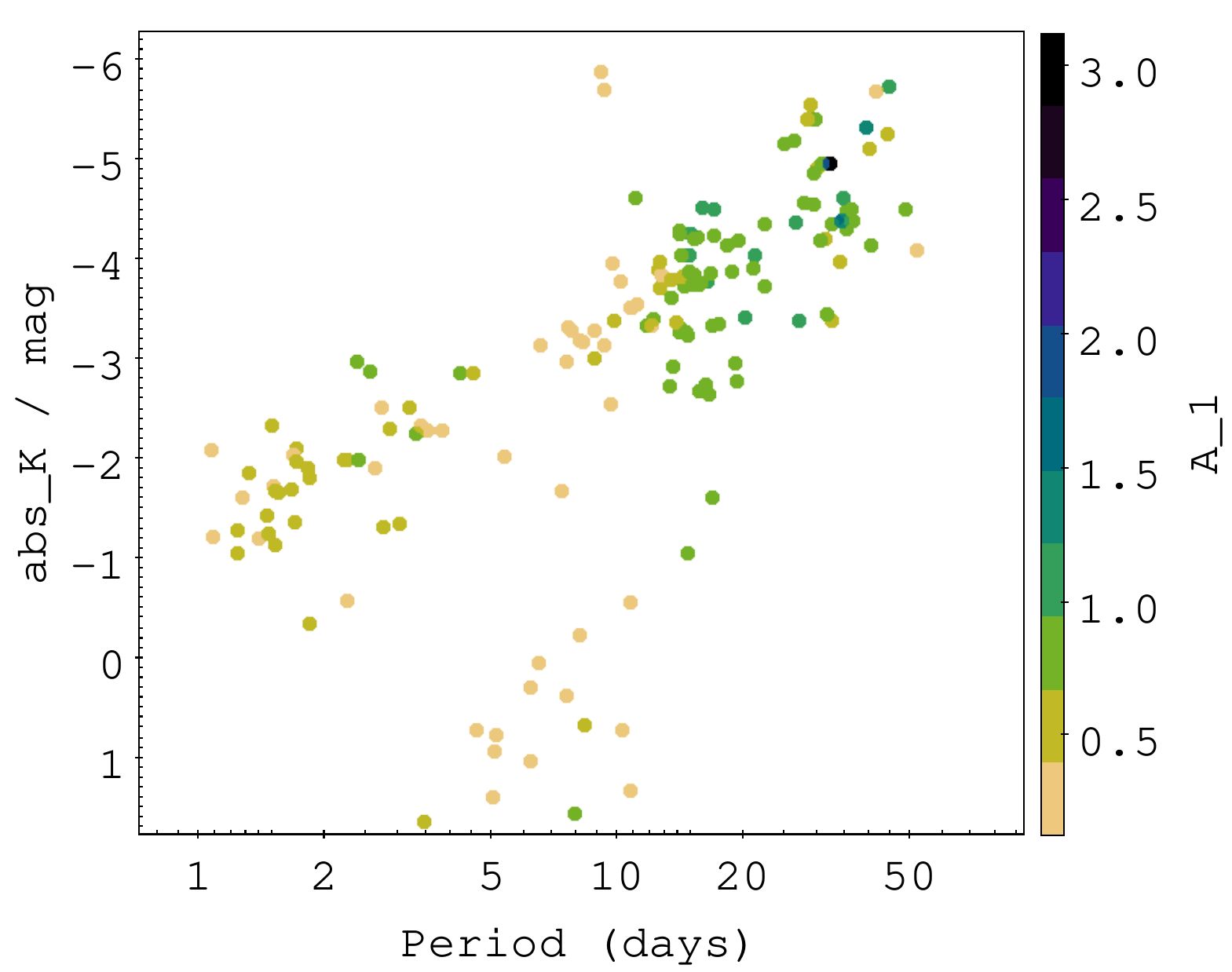}

  \caption{An attempted Period luminosity plot of OGLE type-II cepheids. OGLE amplitudes are also shown in the colour map (see text description). A sample of stars is seen to deviate from the linear P-L law in a fork like pattern. A better picture of this plot with lower scatter is likely to emerge with newer GAIA data releases and from surveys such as VVV. }
  \label{fig:PLlaw}
\end{figure}

\begin{figure}
    \centering
    \includegraphics[width=0.49\textwidth]{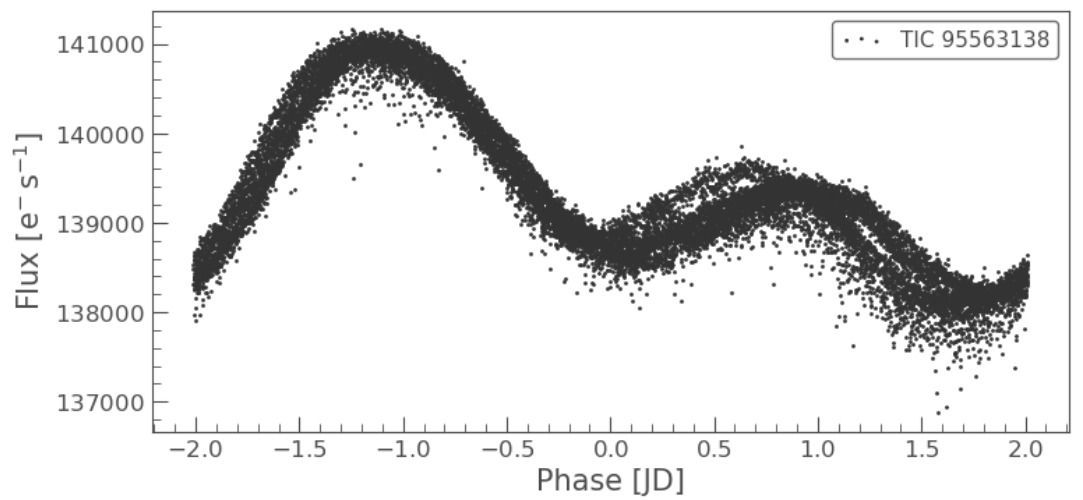}
    \caption{KTCOM lightcurve collected from TESS archive: KT COM has been listed in the GCVS catalogue as a suspected Type-II cepheid. The lightcurve from TESS data shows a characteristic secondary bump in the lightcurve, however, the secondary peak seems to have a variable amplitude between cycles.}
    \label{fig:ktcom}
\end{figure}
\begin{figure}
		\centering

		\includegraphics[width=0.49\textwidth]{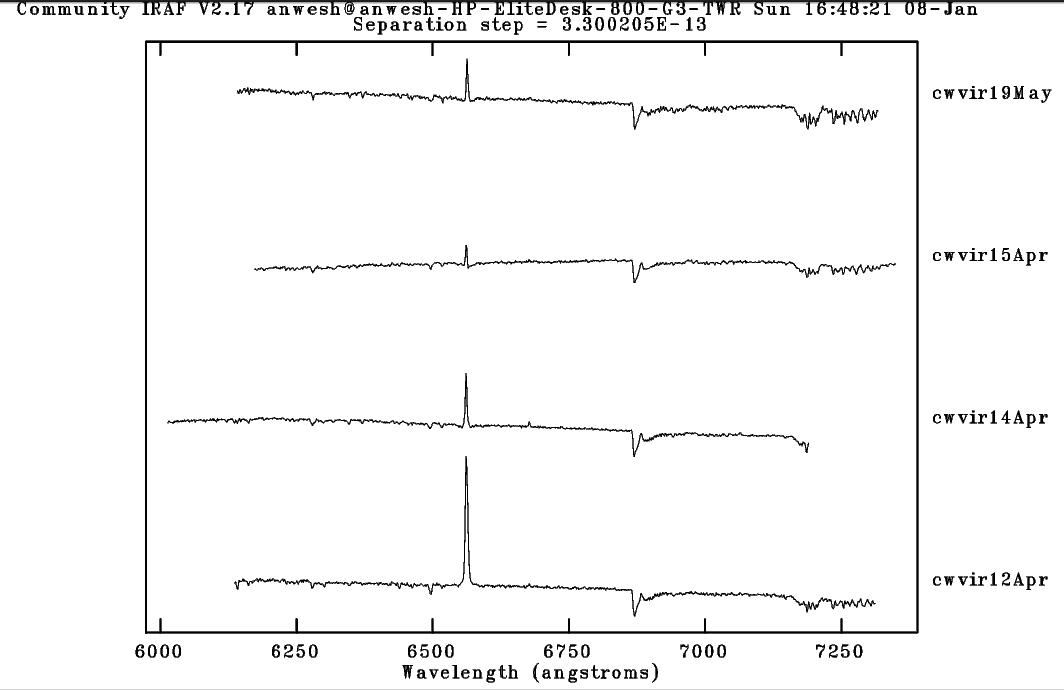}
		 
		\caption{ Variation in emission lines of type-II cepheid W VIR : spectra of W VIR collected over 4 epochs is shown. Variability of the emission line is seen as well as a reddening of the spectra is observed coincident with a reduction in emission line strength. }
      
		\label{fig:WVIR}
\end{figure}

Similarly spectroscopic followup to determine parameters such as metallicity, temperature and evolutionary history of these sources is also important. Dedicated observing runs focusing on the numerous variable stars being discovered is one of the areas where small and medium class telescopes can contribute significantly. Facilities such as OGLE (1.3 Meter, \cite{udalski2015ogle}) and TESS ( $\sim$105 mm, \cite{chrisp2015optical}) have demonstrated excellent use of "small" aperture telescopes. 1-2 meter class telescopes should be capable enough to follow up on sources discovered in these surveys. In particular phase resolved spectroscopy and polarimetric studies are one area where small and medium telescopes can provide useful and significant insights. These observations are likely to be telescope time intensive; highlighting the necessity of specialised observing instruments that have high throughput and are low maintenance.

\section{Type-II cepheids: a case in point}

Type-II cepheids are pulsating variable stars that are usually fainter by 2-3 magnitudes when compared to classical cepheids of same period. These stars occupy a region slightly below classical cepheids in the H-R diagram. These are typically evolved population-II stars in AGB or post-AGB phases of evolution. These stars were identified first by \cite{baade1958problems} as a separate category of evolved population II stars. The GCVS catalogue lists about 300 such stars with a subclassification of CWA (period $>$ 8 days) and CWB ($<$ 8 days) type. With the advent of OGLE survey, and discovery of a large number of these sources, a better subclassification has been suggested by \cite{soszynski2008optical} to split these stars into three separate categories as BL Her ($<$ 4 days period), W Virginis (between 4 to 20 days) and RV Tau ($>$ 20 days) types. This classification has physical significance as the period histogram of type-II cepheids typically show minima at periods close to 4 and 20 days (\cite{bono2020evolutionary}).

Although fainter than classical cepheids, type-II cepheids are still significant distant indicators for regions that contain only older stellar populations (e.g. Dwarf spheroid galaxies and globular clusters \cite{braga2018structure} ) where classical cepheids are absent. \cite{wielgorski2022absolute} have presented P-L laws for shorter period (which are usually more useful as distance indicators) sample of type-II cepheids. A similar analysis is shown in figure \ref{fig:PLlaw} except we have included longer period variables also. The aim here is to probe characteristics these stars themselves. The sources are from OGLE database and were crossmatched with that of GAIA EDR3 (\cite{collaboration2016description}, \cite{brown2021gaia}).  The sample was filtered to include only sources that have \texttt{RPLX} $> 6$. This parameter is indicative of confidence levels of the parallax measurement (\texttt{PLX/e\_PLX}) and is expected to improve with later releases of GAIA. The sources were cross matched with 2MASS (\cite{skrutskie2006two}) for K magnitudes and then the absolute magnitudes were calculated from GAIA parallax. The pulsation amplitude of OGLE is also included as a colour-map in the plot. The plot shows a fainter population of small amplitude stars trailing behind in a fork like pattern. A better picture of this plot with lower scatter is likely to emerge with newer GAIA data releases and from surveys such as VVV \footnote{https://vvvsurvey.org/}. These and other peculiar objects seen in the P-L plot are prime candidates for spectroscopic followup. A similar interesting case for followup is that of KT COM, this source has been listed as a suspected  variable  in the GCVS catalogue. The lightcurve from  TESS is shown in figure \ref{fig:ktcom}. The star shows characteristic secondary bump of short period type-II cepheids. However, the secondary peak seems to have variable amplitude in-between different cycles. This star has period of about 4 days which lies at the transition point between BL Her and W Vir type stars and is fairly bright ($V_{mag} 8.23 $) making it a good candidate for follow-up observations.

Intermediate to long period type-II cepheids also present unique and interesting aspects for further study. Spectra of W Virginis (observed using the MFOSC-P instrument \cite{srivastava2021design}) on four different nights are shown in \ref{fig:WVIR}. This star seems to show a variable emission in H-$\alpha$. Of particular interest is the reddening of the spectrum along with a concurrent diminishing of emission line strength. This could be a signature of infrared excess caused by either episodic or pulsation-driven mass loss. It will be interesting if this can be probed separately by means of phase resolved polarimetry or SWIR photometry.

Considering these interesting science cases, we propose for dedicated observing instruments from various facilities of the Mt. Abu observatory. The instruments discussed focus on observing pulsating variables (in particular type-II cepheids) in phase resolved spectroscopy and Visible/ SWIR photometry.

\begin{figure*}
		\centering

         \begin{subfigure}{0.59\textwidth}
		\includegraphics[width=0.99\textwidth]{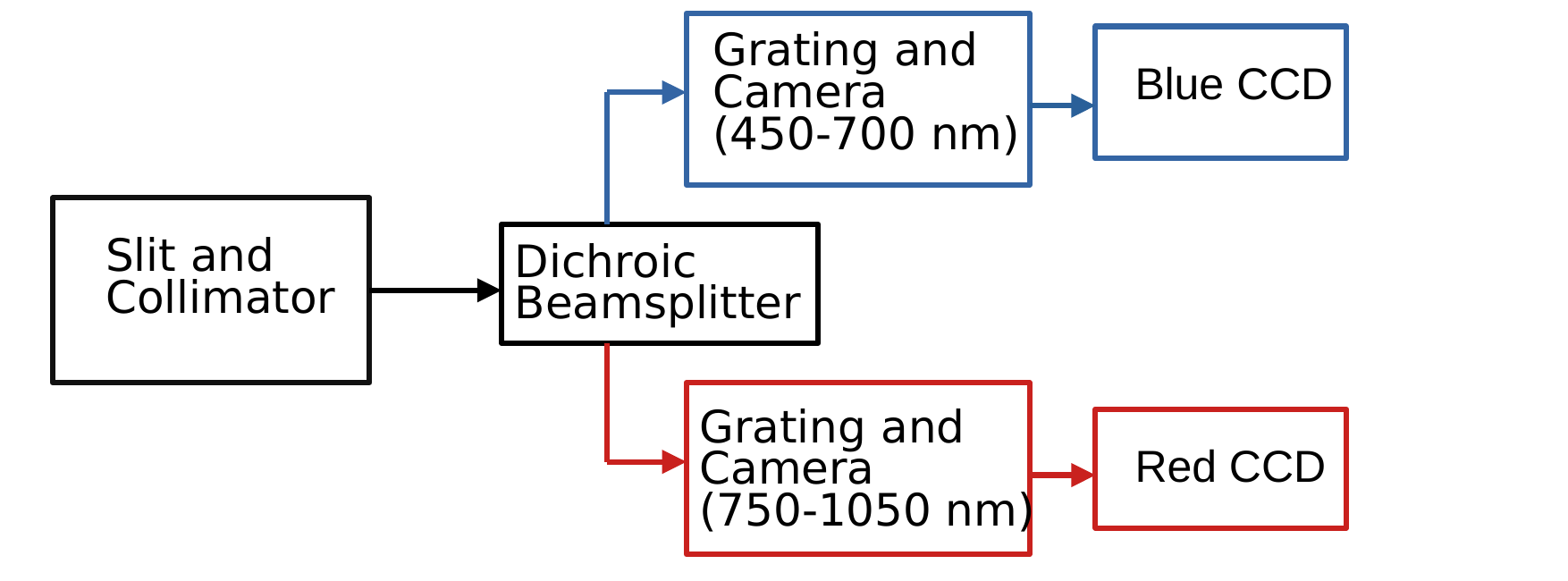}
			\caption{Block diagram }\label{fig:spect_block}
		\end{subfigure}
        
       \bigskip          
        \begin{subfigure}{0.75\textwidth}
		\includegraphics[width=0.85\textwidth]{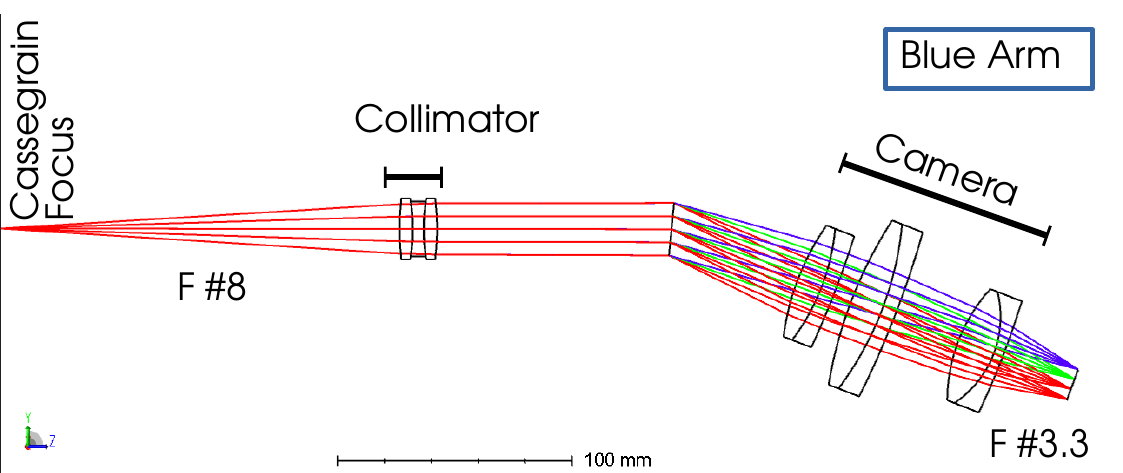}
			\caption{Blue channel }\label{fig:spect_layoutvis}
		\end{subfigure}
		\bigskip
		\begin{subfigure}{0.75\textwidth}
			
       \includegraphics[width=0.85\textwidth]{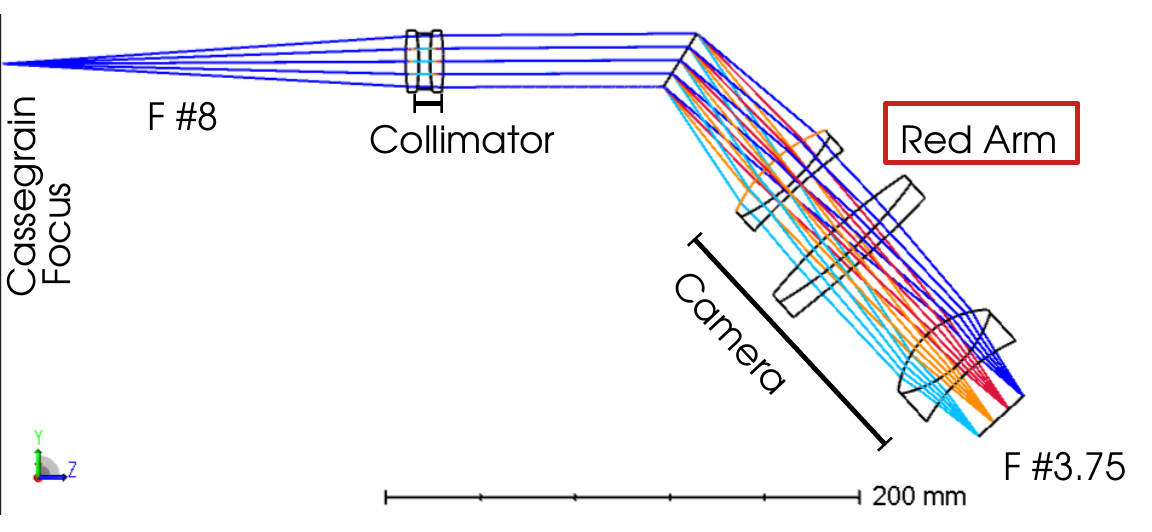}
			\caption{Red channel}\label{fig:spect_layoutnir}
		\end{subfigure}

		\caption{ The proposed high throughput two channel spectrograph:
          The block diagram of the instrument is shown in (a). The spectrograph is split into two separate channels for blue and red wavelength range so that in each channel the components are optimised for high throughput. The slit for the spectrograph is at the cassegrain focus of the 2.5 M telescope. Standard collimator camera type design with a grating in the collimated beam section is followed for both channels. The collimator optic is common for both the channels; consisting of a Hastings triplet. The splitting between the channels is done after the collimator using a dichroic beamsplitter such as Newport 20CMS-45 or Edmund optics 64-451 centered at around 700nm. The camera is implemented using off-the-shelf doublets from Edmund optics and Thorlabs. The layout of the blue channel is shown in (b) and that of the red channel is shown in (c). The complete optical chain including the VPH gratings (described in figure \ref{fig:efficiency}) is implemented using off-the-shelf components.}

		\label{fig:cam}
  \end{figure*}

\begin{figure*}
		\centering
		
    \begin{subfigure}{0.85\textwidth}
		\includegraphics[width=0.9\textwidth]{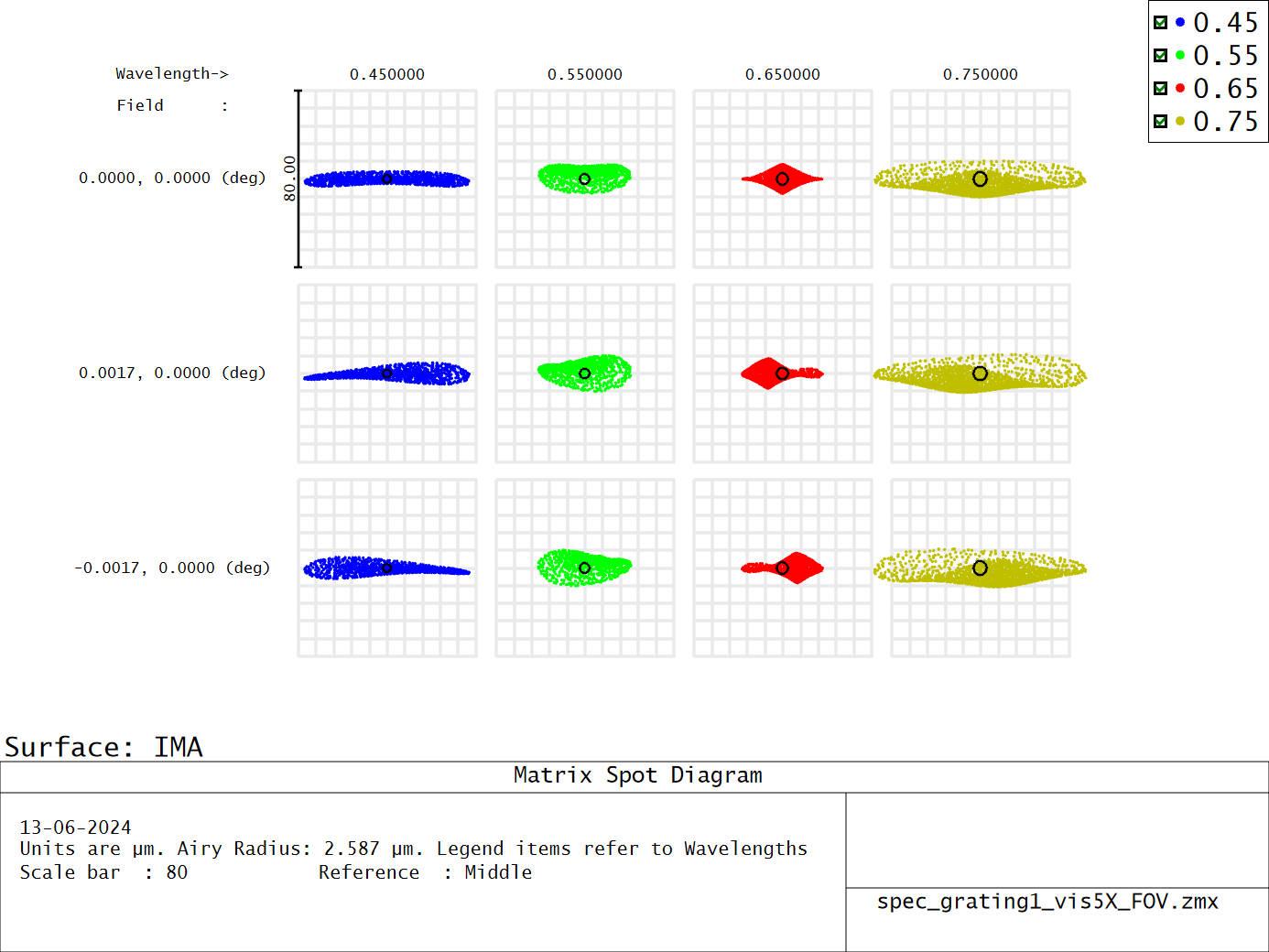}
			\caption{Spot diagram of the blue channel}\label{fig:visspot}
		\end{subfigure}
		\bigskip
		\begin{subfigure}{0.85\textwidth}
			
       \includegraphics[width=0.9\textwidth]{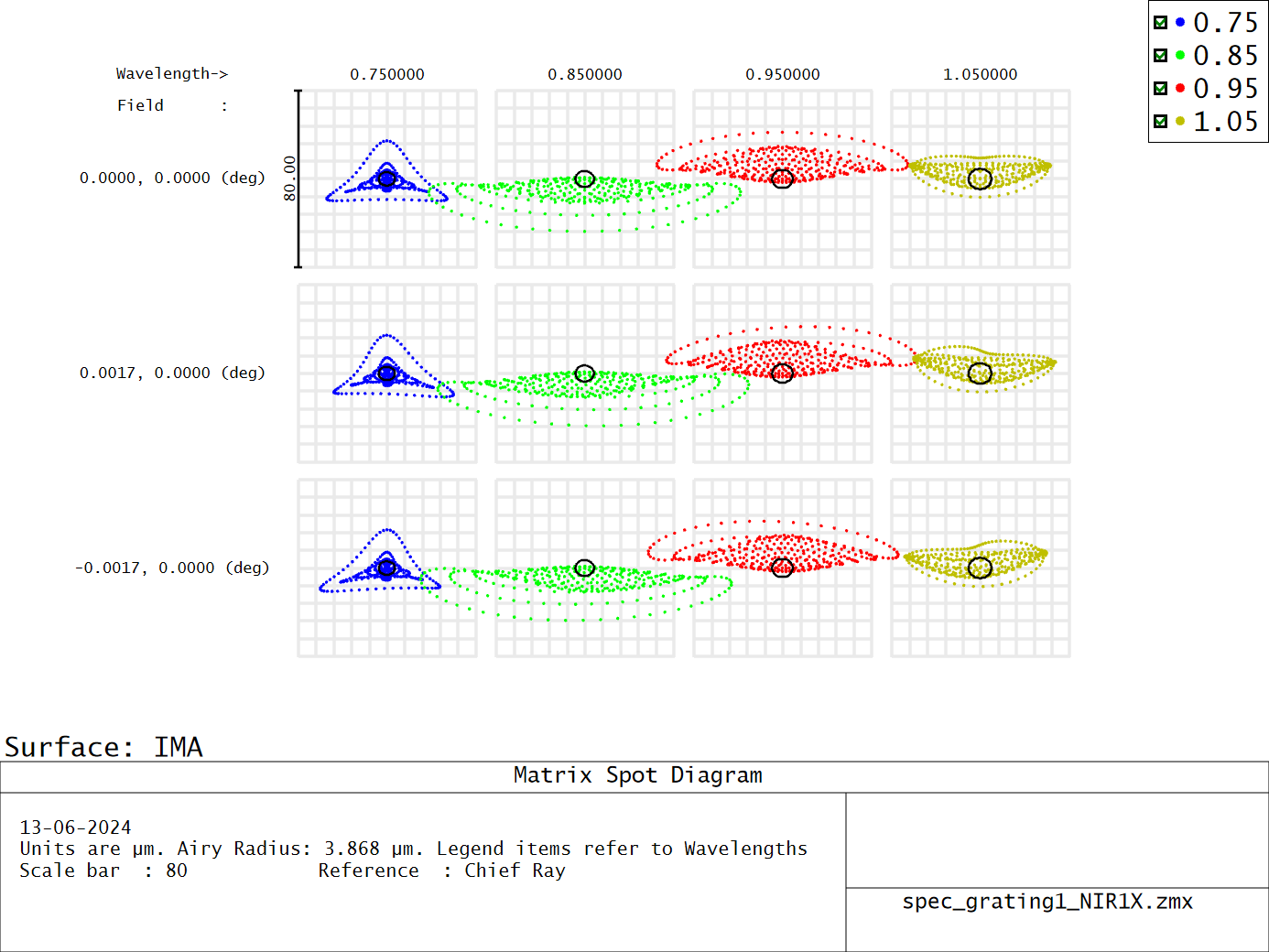}
			\caption{Spot diagarm of the red channel}\label{fig:NIRspot}
		\end{subfigure}

		\caption{Matrix spot diagram of the spectrograph: The spot of the blue and the red channel are presented in (a) and (b) respectively.  During optimisation of the optical performance, the spot was allowed to be larger in the direction normal to the dispersion axis (i.e. along the slit) so that the best spectral resolution is achieved along the direction of dispersion. The "spot width" is aimed to be within 1 pixel for the blue channel and within 2 pixels for the red channel.(details of the CCD arrays are provided in table \ref{tab:spect})  The resolution of the spectrograph is expected to be pixel size limited for a total field of +/- 6 arcseconds and is shown to be reasonably uniform along the slit.} 
		\label{fig:specspot}
\end{figure*}

\section{Proposed instruments for the Mt.
Abu Observatory:}

The Mount Abu observatory is a facility initiated and operated by Physical Research Laboratory, Ahmedabad since 1994. Presently the site is home to a number of telescopes including the 1.2 meter telescope (\cite{banerjee1997effect}), the newly installed 2.5 Meter telescope (\cite{chakraborty2024prl}), as well as a 50 (\cite{ganesh2013automated}) and a 43 cm telescope. The site is located at an altitude of 1680 meters above sea level and enjoys a large number of observing nights and fairly low water vapour content\footnote{https://www.prl.res.in/~miro/}. A brief analysis of infrared capabilities of the site  has been discussed in \cite{prajapati2023near} . The site is definitely among one of the better sites available to Indian astronomers.

We plan for the following instruments to observe pulsating variables from the facilities of Mt. Abu observatory. These instruments are to augment the existing capabilities of the observatory such as the photo-polarimter (\cite{deshpande1985astronomical}) and EMPOL (\cite{ganesh2020empol}).

\begin{itemize}
    \item A two channel faint object spectrograph for 1.2 and 2.5 meter telescope
    \item SWIR imager based on InGaAs array for the 1.2 meter telescope
    
    \item Wide-Field camera and SWIR camera for the 50 CM telescope
\end{itemize}

The common design goal for all of the above instruments is to achieve both high throughput as well as ease of implementation by keeping the optical design simple.  We have made use of off-the-shelf optical components whenever possible to reduce the cost and time for implementation. We present the motivation, optical design and expected performance and observational scope of these instruments.

\begin{figure*}
    \includegraphics[width=0.85\textwidth]{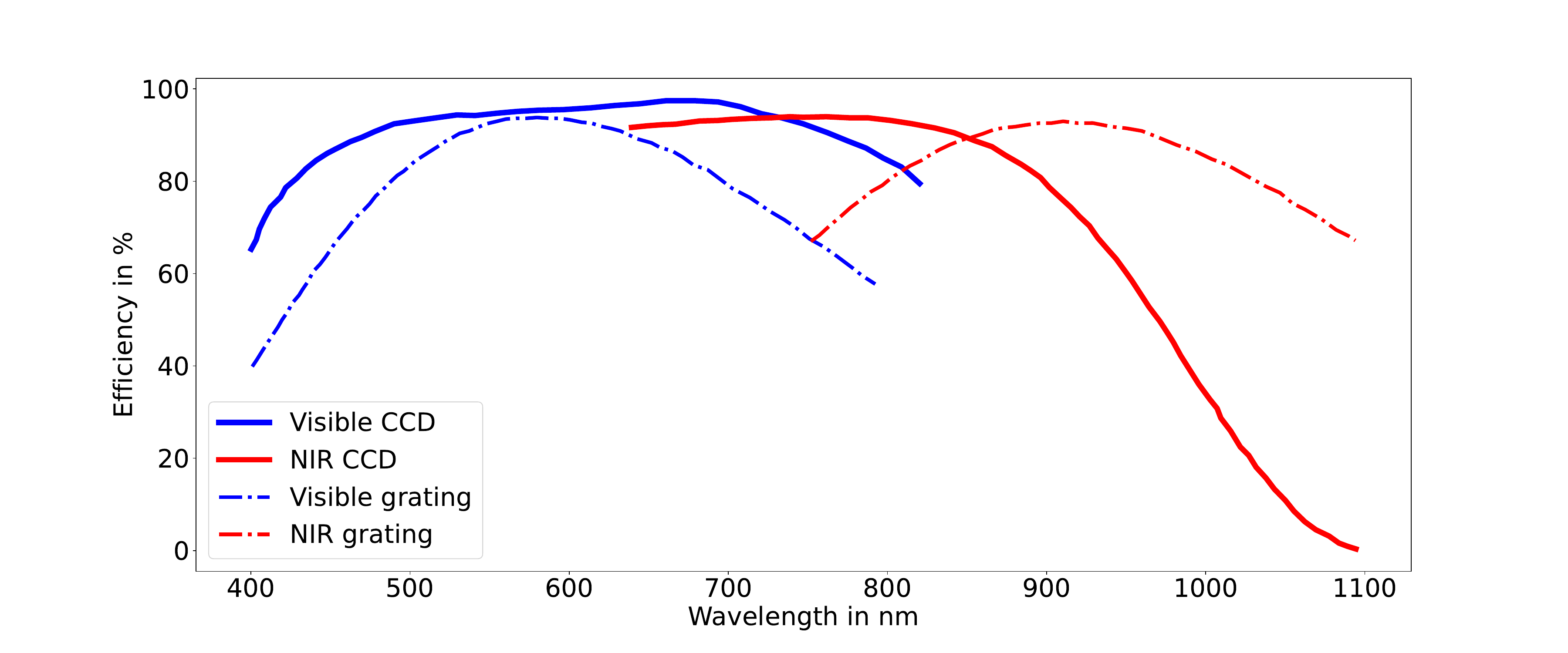}
			\caption{Efficiency as a function of wavelength for critical components of the spectrograph. The VPH gratings achieve fairly high diffraction efficiency  at their design wavelengths. Curves for the VPH grating GP3506G for the blue channel and VPH grating GP3510M for the red channel is shown. Both of which are off-the-shelf components from Thorlabs. Quantum efficiency curves of suitable CCD sensors are also shown. Pixis 100B\_eX CCD from Princeton instruments is found to be best suitable for the blue channel and the DU920PBEX2-DD from Andor is found to be most suitable for the red channel. Both CCDs are in spectroscopy optimised formats (table \ref{tab:spect}). }\label{fig:efficiency}
\end{figure*}

\subsection{A two channel faint object spectrograph for the 1.2 / 2.5 meter telescopes }

A spectrograph operating 450-1000nm range is planned to observe phase resolved spectra of pulsating variables (particularly type-II cepheids). The spectrograph needs to have good optical throughput to conduct followup observation on new type-II cepheids being discovered. Alongside, it is also desired that the spectrograph is built from off-the-shelf components; this is necessary for a quick turnaround time of the instrument as well as to reduce the total cost.  The spectrograph is designed to have two separate channels for blue and red wavelength ranges and is planned to have no moving components to achieve high reliability in operation. The spectrograph will be suitable for observation on both the 2.5 meter telescope as well as the 1.2 meter telescope. The following are the defining features of the spectrograph:  \\

\begin{table}[htb]
\tabularfont
\caption{Specifications of the two channel spectrograph. The spectrograph aims to achieve high throughput while using mostly off-the-shelf components. }\label{tab:spect} 
\begin{tabular}{lcc}
				
				\topline
				\textbf{} & \textbf{Blue channel} & \textbf{Red channel} \\
				\hline
                
                 Wavelength & 400-720 nm & 750-1090 nm\\
				
                Slit-width & 50$\mu$m(0.5" on 2.5M ) &  \\
                   & 75$\mu$m (1" on 1.2M) & \\
                Grating make & Thorlabs & Thorlabs \\
                Grating type & Transmitting & Transmitting\\
                Grating Model & VPH GP3506G & VPH GP3510M\\
                 Collimator & Hastings Triplet & Hastings Triplet \\
                
                 Camera F\#& 3.3 &3.75\\
              
                 Throughput & 30\% & 30\%\\
                
                 CCD make & Princeton-Pixis & Andor-Newton \\
                 Model & 100B\_eX & DU920PBEX2-DD \\
                 Format & 1340 by 100 & 1024 by 256 \\
                 Pixel size & 20 micron  & 26 micron \\
                 Resolution & 0.3nm /pixel & 0.42nm /pixel\\
                 Read noise & $3 e^- /pixel $& $4 e^- /pixel $ \\
                 Dark current &$ 0.001 e^- / p/S $ & $ 0.003 e^- / p/S $\\
                
                \hline
 			\end{tabular}

\end{table}

\begin{itemize}
    \item A two channel design that  splits 400-1000nm wavelength range using a high efficiency \textbf{dichroic beamsplitter}.
    \item Using \textbf{VPH gratings} for high diffraction efficiency (\cite{barden2000volume}, \cite{monnet2002volume}) over wide wavelength range.
    \item  Use of separate blue and red optimised \textbf{spectroscopic format CCD}; each with high QE in its respective wavelength range
    \footnote{https://andor.oxinst.com/} , \footnote{https://www.princetoninstruments.com/}
    \item Off-the-shelf \textbf{achromatic doublet and triplet  lenses} 
     
\end{itemize}

The block diagram of the design is shown in figure \ref{fig:spect_block} The implementation requires a dichroic filter such as Newport\footnote{https://www.newport.com/} 20CMS-45 or Edmund Optics\footnote{https://www.edmundoptics.in/} 64-451 to split the blue and red channels without losing efficiency. The special feature of these dichroics is that of good efficiency in both transmission and reflection as well as a sharp ($\sim$ 50 nm) transition. The optical layout of the blue and red channel are shown in figure \ref{fig:spect_layoutvis} and \ref{fig:spect_layoutnir}. As shown in the figure, the slit and the collimator is common for both channels. The Edmund optics hasting triplet (optimised for infinite conjugate imaging) is selected as suitable for this purpose. The camera section is implemented using off-the-shelf doublets from Edmund Optics for both the channels. Both channels use off-the-shelf VPH gratings as well as spectroscopy optimised CCDs to maximise throughput. The red channel makes use of deep depletion type CCD from ANDOR which minimises fringing effects. CCD parameters such as dark and read noise of the CCDs are summarised in table \ref{tab:spect}. Details on the diffraction efficiency of the VPH gratings and the CCDs is described in detail in figure \ref{fig:efficiency}. The spectrograph is designed to have about 0.3 nm/pixel (3$\AA$ per pixel ) resolution per pixel in the blue channel and 0.42 nm/pixel (4.2$\AA$ per pixel) in the red channel. The optical performance of the spectrograph  is shown in figure \ref{fig:specspot}. The spot of the spectrograph was optimised to be smaller along the dispersion axis and allowed to be larger along the slit. This lets us achieve for a strictly pixelsize limited resolution for both the channels. The spectrograph will also be usable on the F\# 13 beam of the 1.2 meter telescope without vignetting.  \\

\begin{figure}
  \centering
 
		\begin{subfigure}{0.49\textwidth}
         \includegraphics[width=0.99\textwidth]{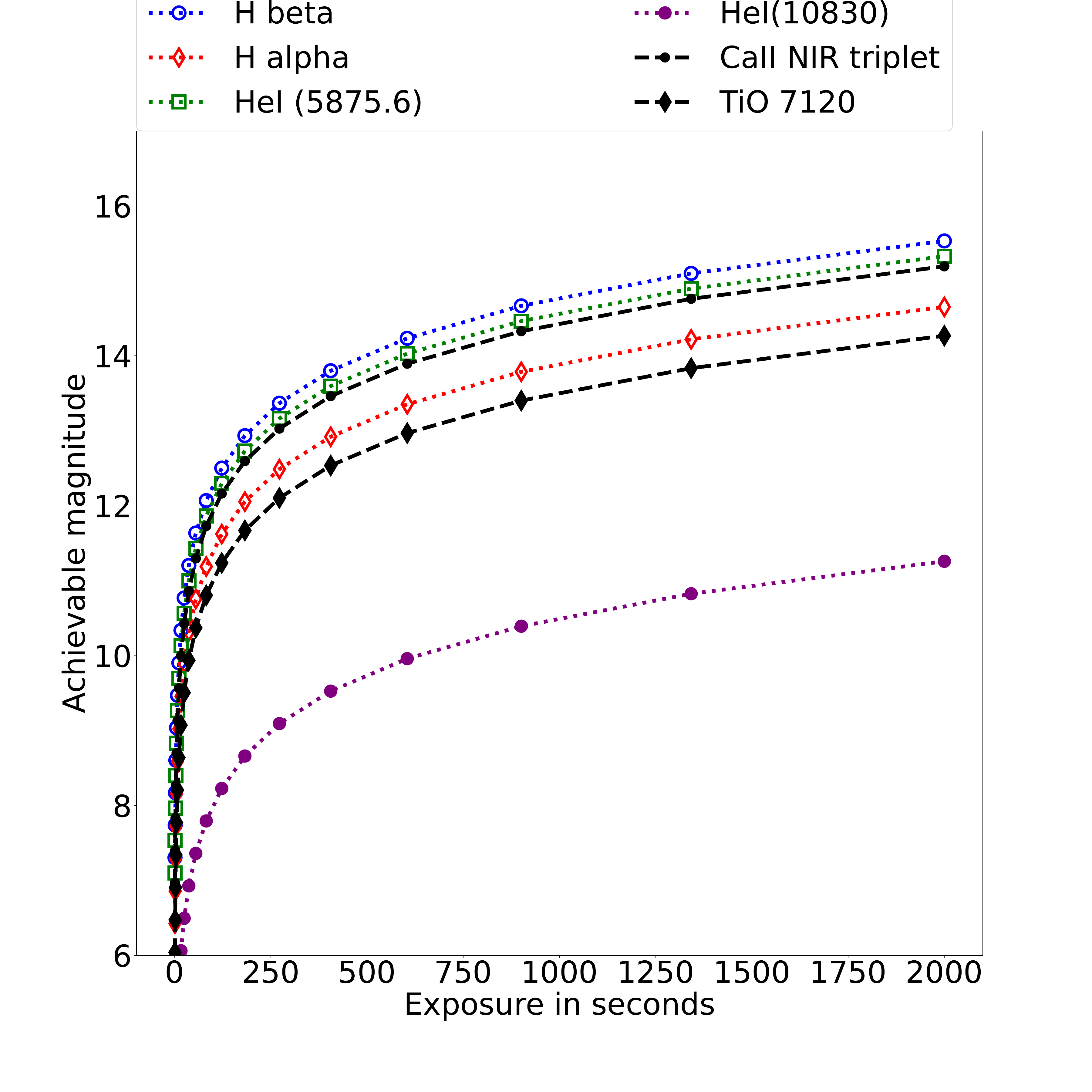}
			\caption{Expected magnitudes for different wavelengths}\label{fig:magspectrum}
		\end{subfigure}
        \bigskip
         \begin{subfigure}{0.49\textwidth}
        \includegraphics[width=0.99\textwidth]{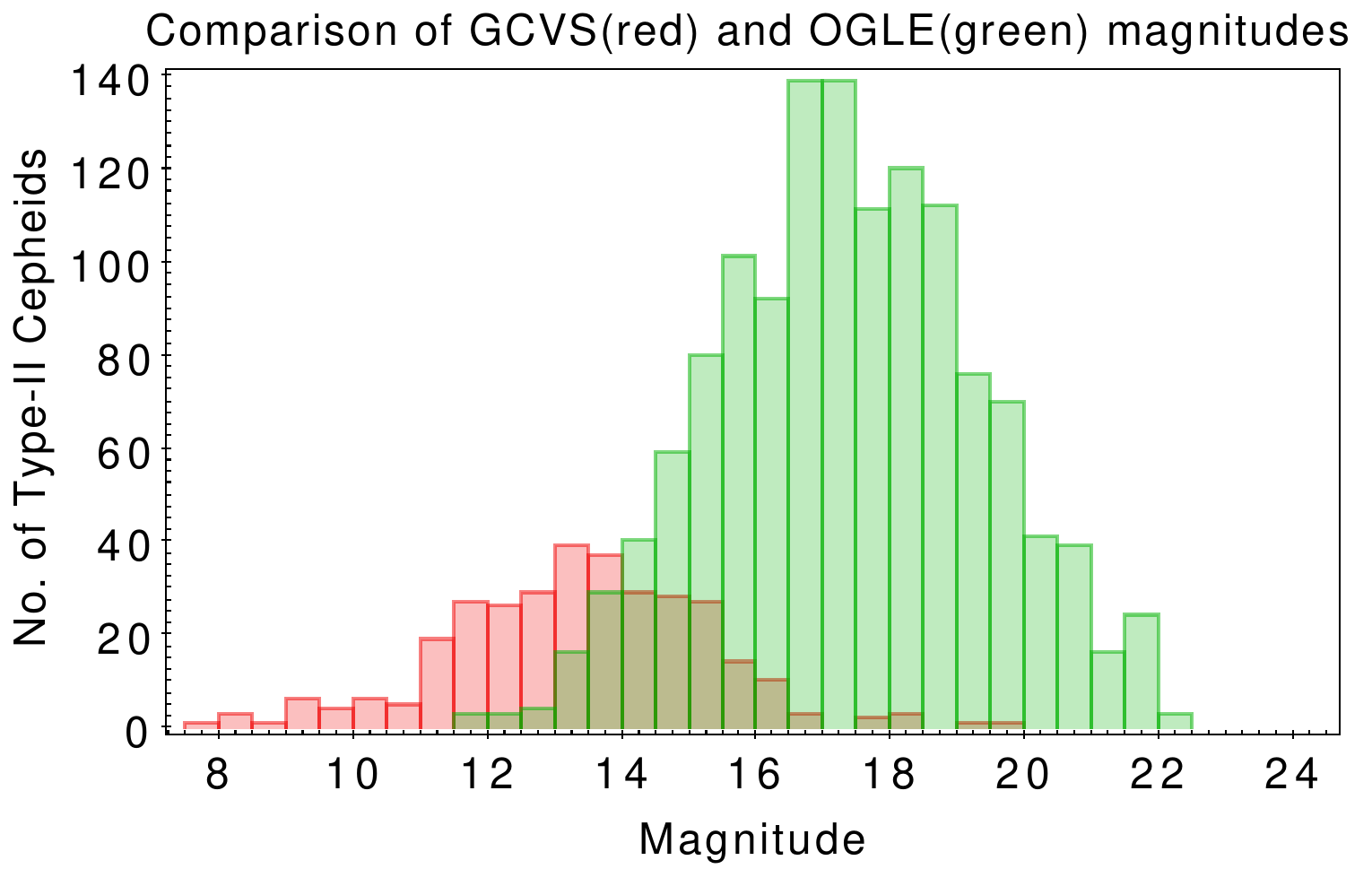}
			\caption{}\label{fig:OGLEVspec}
		\end{subfigure}
		\bigskip
		\begin{subfigure}{0.49\textwidth}
		\includegraphics[width=0.99\textwidth]{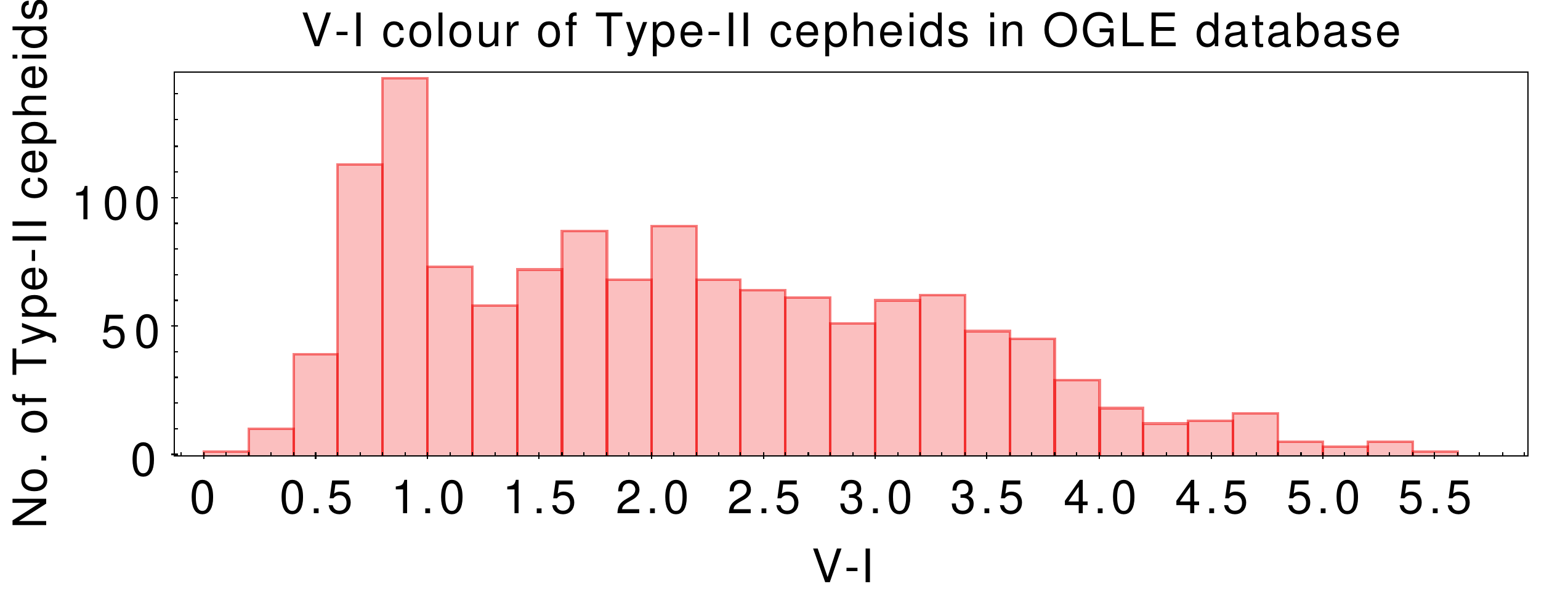}
			\caption{}\label{fig:OGLEV_I}
		\end{subfigure}

   \caption{Achievable magnitude limits v exposure time for various wavelength regions is shown in (a) for the two channel spectrograph. Histogram of observed type-II cepheid magnitudes in GCVS and the OGLE survey is shown in (b). V-I colour distribution in the OGLE survey is shown in (c) to illustrate that these stars are typically brighter in longer wavelengths as they have low surface temperatures.}
   \label{fig:magspectrumT}  
\end{figure}

Compared to instruments that combine an imager as well as spectrograph, this implementation has the advantages of making use of cheaper spectroscopy optimised CCD formats, being more reliable (without any moving parts) and better throughput with only off-the-shelf lenses. The proposed design offers a few appealing features for observing pulsating variables. The blue channel covers both H-$\alpha$ and H-$\beta$ regions with resolution $>$1000 and fairly high throughput. The red channel covers important spectral features for characterising late type stars ( \cite{white1978photoelectric}). Of particular interest is the HeI line at 10830 $\AA$. This line is indicative of chromospheric activity(\cite{vaughan1968helium}) in late type stars. \cite{schmidt2004spectra} have reported that this line appears in emission (along with H$\alpha$ and H$\beta$) for type-II cepheids and shows variability with pulsation phase.  The HeI line at 5876 $\AA$ (\cite{lanzafame1995helium}) is sometimes used as a proxy for the HeI 10830 $\AA$ line. The 5876 $\AA$ HeI line is typically not resolved at low resolution spectra and may only be observed if it shows up as an emission line. The CN line at 8120 $\AA$ and TiO at 7120 $\AA$ (\cite{white1978photoelectric}) are very useful for characterising long period and highly evolved variable stars such as RV Tau and Mira variables. CaII K lines at 3934 \AA have been known to show emission profiles for classical cepheids (\cite{kraft1957calpha}). The CaII triplet in the NIR (8498 $\AA$, 8542 $\AA$ and 8662 $\AA$) along with H$\alpha$ profiles have been used for studying the pulsational properties of these stars(\cite{hocde2020pulsating}). \\

The expected magnitude V exposure time plot for different spectral wavelength ranges is presented in figure \ref{fig:magspectrum} for different spectral ranges (for 2.5 meter use). The magnitude limits are for fairly high SNR($\approx$100) and are considered to be photon noise limited as the dark and read noise of the CCDs are fairly low( table \ref{tab:spect}). The method of calculating the same is discussed in appendix-A. The V band magnitude distribution of type-II cepheids in GCVS and in OGLE survey is  presented in figure \ref{fig:OGLEVspec} for comparison. The blue channel is expected to be able to follow up most of GCVS sources and a significant sample of OGLE sources. The same goes for lines such as CN  and CaII triplets in the red channel. The red channel unfortunately suffers from the limit in CCD QE after 950 nm and will be able to follow stars only up to 10th magnitudes in lines such as HeI. However, it is also of note that most pulsating variables are  typically brighter at 1000 nm than in the V band. For example, V-I colour of type-II cepheids in the OGLE survey is shown in figure  \ref{fig:OGLEV_I}. This is expected since these stars are evolved stars and have a low surface temperature. The Red channel of the proposed spectrograph is likely to provide useful insights in the study of these type of stars. The spectrograph will focus more on the fainter limit of the OGLE survey when used along with the 2.5 meter telescope and will focus on obtaining finely phase resolved spectra of brighter sources when operated on the 1.2 meter telescope. The platescale for the 2.5M telescope is 10.3” per mm and for the the 1.2 M telescope it is about 13.2” per mm. Accordingly, the recommended slit sizes of 50 and 75 microns cover about 0.5” on sky for 2.5M and 0.97” for 1.2M respectively. A larger slit on the 1.2 M has the benefits of offsetting some amount of light capture disadvantage of the smaller telescope. This comes at a loss of resolution of about 20-30 percent depending on the wavelength range. If higher resolution is desired for an observing program focusing on brighter stars, then the smaller slit can also be used on the 1.2 M telescope.\\

\begin{figure*}
\centering
    \includegraphics[width=0.85\textwidth]{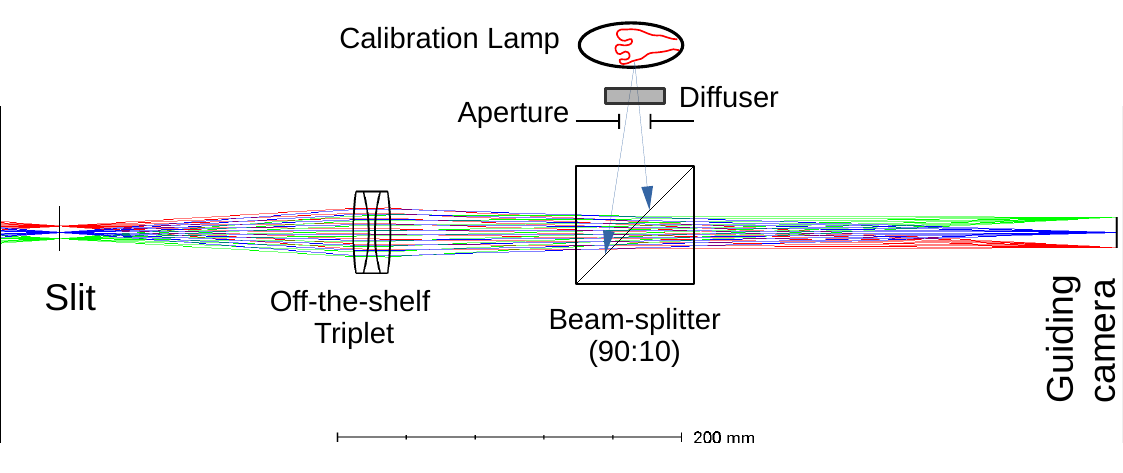}
			\caption{Layout of the guiding camera for the 2.5 M telescope: the guider reimages a field of about one arcminute centred on the slit onto a guiding camera. The system is realised by an off-the-shelf triplet. The triplet also serves to deliver illumination from the calibration lamp back onto the slit. A beamsplitter (90\% transmitting 10\% reflective such as Thorlabs BSN10) maybe used to combine both the functionalities.}\label{fig:guiderlay}
\end{figure*}

\begin{figure}
  \centering
 
         
        \includegraphics[width=0.99\columnwidth]{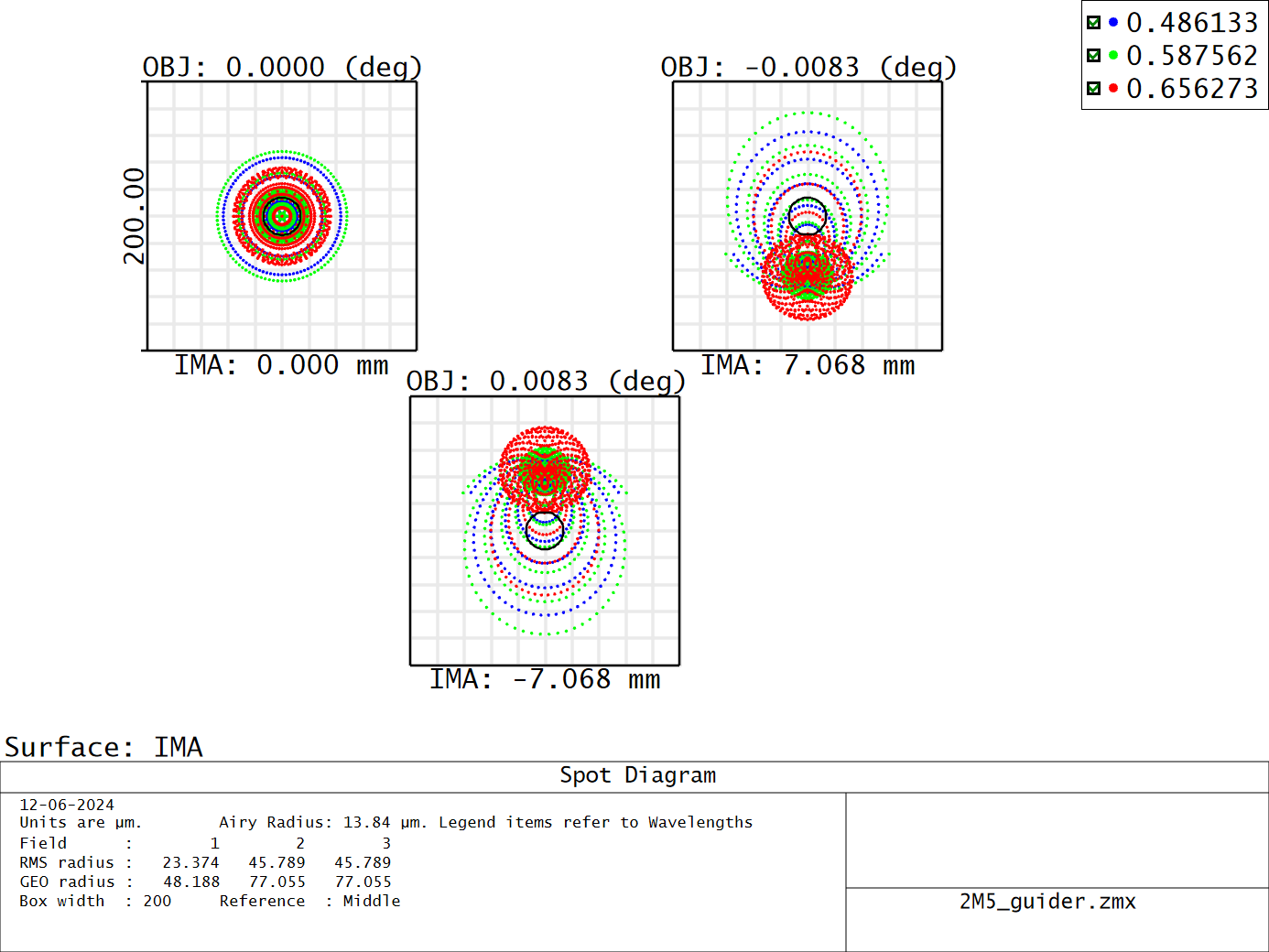}
			\caption{The spot diagram of the guiding camera: the guiding camera has fairly good spot over one arcminute field of view. The illustrative box represent about one arc-second. The output of the guiding optics is made intentionally slow (about F\#19) to oversample the seeing disk for accurate guiding. Due to the slower beam a large pixelsize camera (or a camera with larger binning capability) will be desired.}\label{fig:guiderspot}


\end{figure}

Keeping in tune with the ``zero-moving-parts" ideology a combined slit-viewer as well as lamp illumination scheme is presented in figure \ref{fig:guiderlay} and its spot performance is shown in figure \ref{fig:guiderspot}. The optics is realised by a simple off-the-shelf triplet. The triplet reimages about one arcminute of field centred on the slit onto a guiding camera. The guiding camera is yet to be finalised but a sensor with sensitive area of about 15 mm that can operate at fairly high binning--typically 8$\times$8 or higher--will be required. The same optical chain will also deliver the calibration lamp illumination from the spectral lamps back to the slit. The guide-camera  views the cassegrain focus through a reflective slit. The slit is fixed at  an angle (e.g. \cite{willstrop1976optics}) to allow for “off-axis” viewing of the cassegrain focus through a fixed fold mirror. During preliminary acquisition, the field will be confirmed and the target star positioned onto the slit using the guide camera. (The necessary beam folding arrangement will be decided during opto-mechanical design of the instrument and is not shown in figure \ref{fig:guiderlay}.) \\

Pencil style form factor calibration lamps from Newport are a good match for the form factor of the instrument. The wavelength calibration optics will be the same optical system that is used for the slit-viewer (figure \ref{fig:guiderlay}). Since the lamps are fairly small, multiple lamps can illuminate the diffuser screen which is then reimaged back onto the slit plane by the off-the-shelf triplet. For the blue channel Neon and Argon lamps will be a good match. For the red channel Krypton and Xenon lamps will provide large number of calibration points. The prerogative to select the necessary lamp for their wavelength of interest may be left to the observer.\\

The spectrograph aims to cover the largest spectral region available (while using off-the-shelf spectroscopy optimised CCDs), this is necessary as different variable stars show variability in a range of different spectral lines and good spectral coverage is important to study a range of variable stars. Additionally, this approach also verifies the performance of the optical chain over a larger wavelength region and if a dedicated survey focused on higher resolution is required, this can be achieved using the existing optical system by simply swapping out the existing grating.
As the spectrograph is fairly low resolution, the wavelength solution may only need to be good enough to identify lines. If necessary,  a larger number of calibration lines may be used for the wavelength solution in order to compensate for the sub-Nyquist sampling of the lamp lines and smooth out the random errors from coarse centroiding. \\

 \subsection{ SWIR imager using InGaAs technology}

InGaAs based cameras are an emerging detector technology in the wavelength range of 800-1700 nm. The outstanding feature of these  arrays is that they can be operated with only moderate amount of cooling( \cite{sullivan2013precision}, \cite{sullivan2014near}). These arrays can achieve acceptable levels of dark currents with just thermoelectric cooling. This is in contrast to HgCdTe detectors, which typically have to be cooled below 100K and as such need cryogenic cooling by means of liquid nitrogen, InGaAs detectors can be implemented into an astronomy camera as a much simpler and lightweight instrument (\cite{simcoe2019background};  \cite{batty2022laboratory}). These features make them  attractive for small-medium telescopes where there may be a weight limit and particularly for automated/robotic facilities as there is no need for on-site continuous on-site maintenance to refill liquid nitrogen.  

\begin{figure*}
		\centering
		
        \begin{subfigure}{0.99\textwidth}
			\includegraphics[width=0.95\textwidth]{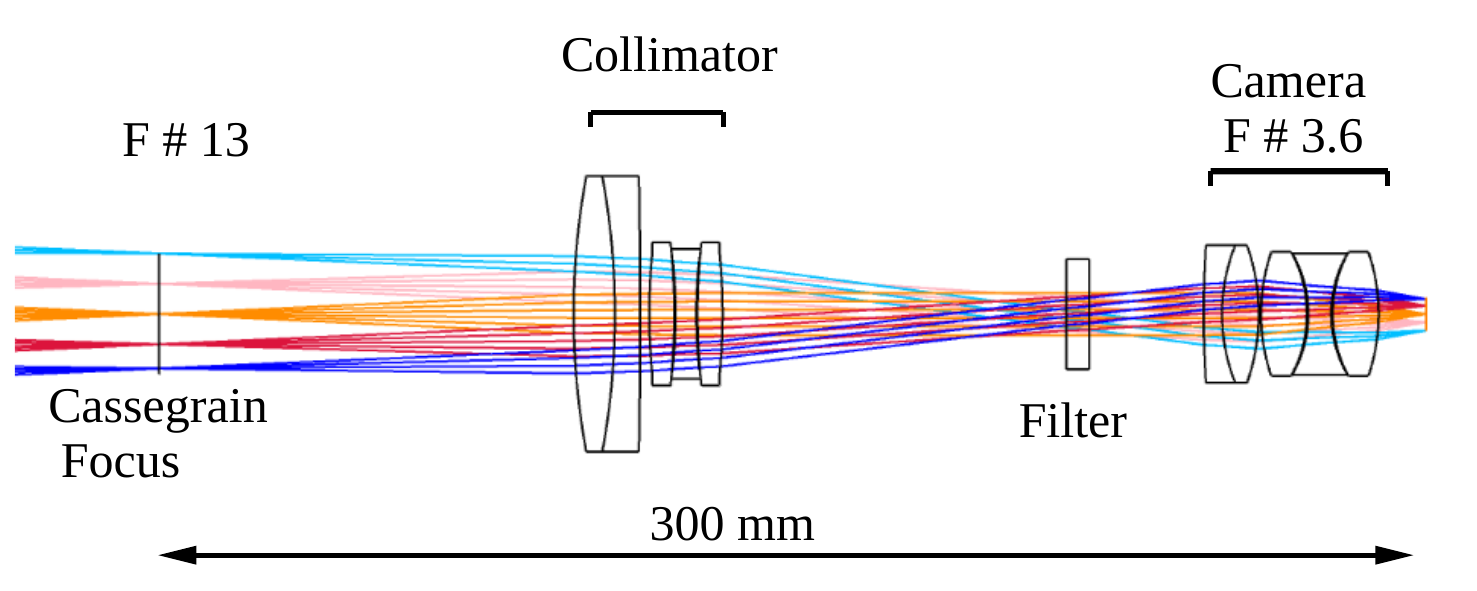}
			\caption{Layout for an SWIR imager for the 1.2M telescope }\label{fig:layoutSony}
		\end{subfigure}
		\bigskip
		\begin{subfigure}{0.99\textwidth}
			
       \includegraphics[width=0.99\textwidth]{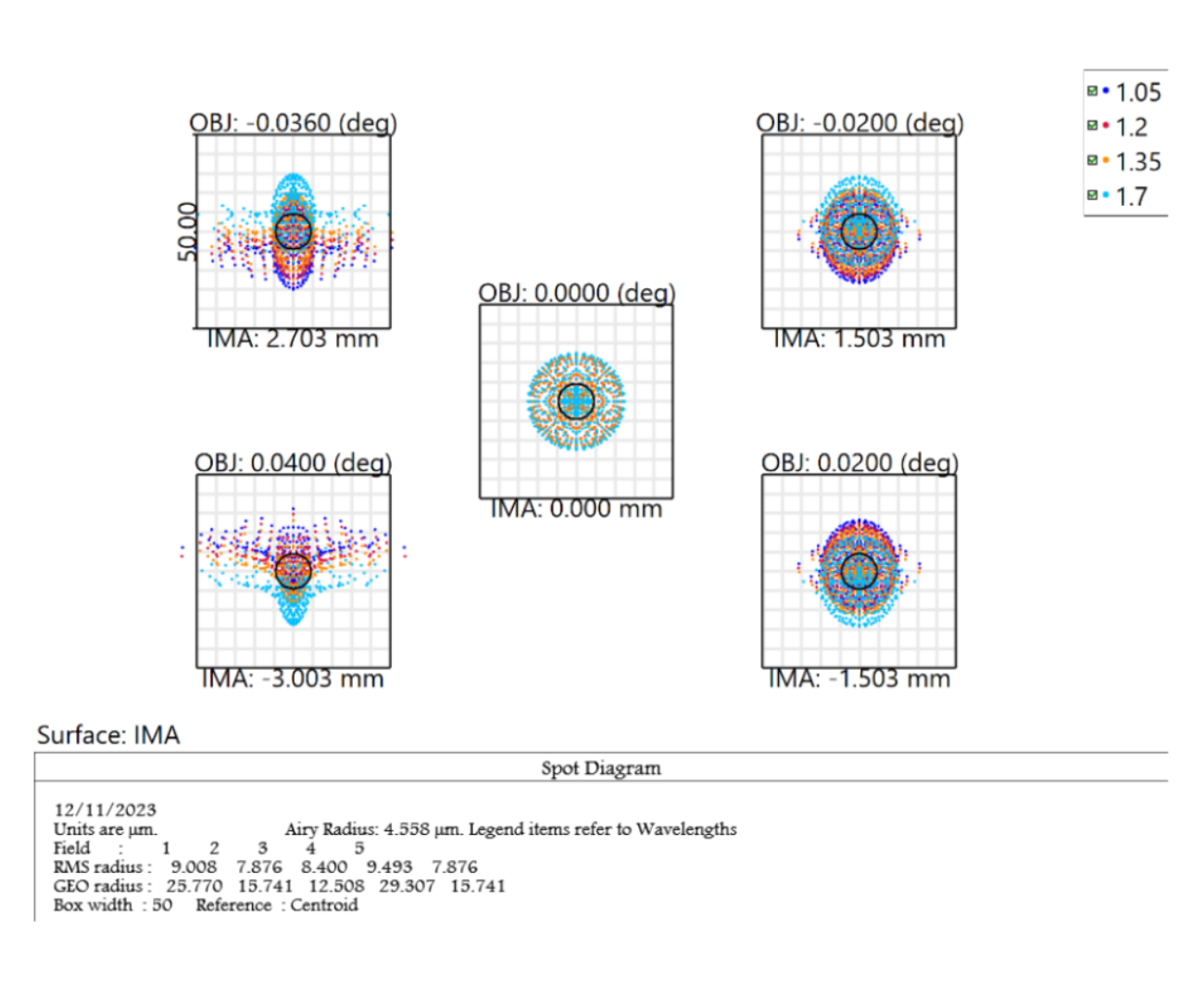}
			\caption{Spot diagram for the imager }\label{fig:spotingaas}
		\end{subfigure}

		\caption{ Optical design and spot performance for an SWIR imager on the 1.2 M telescope: in (a) the layout of the optical design is presented. The design is based on a collimator camera type optics which converts the F\#13 of the telescope focal plane to F\# 3.6 at the detector plane. The fast output of F\#3.6 is necessary to accommodate the small pixel size of the IMX990 sensor. The spot sizes of the design for various field and wavelengths is shown in (b).  The design is implemented using only off-the-shelf lens elements and is expected to perform at the seeing limit for the entire field of view. }
		\label{fig:ingaas}
\end{figure*}

\begin{figure}
		\centering
		
        \begin{subfigure}{0.4\textwidth}
			\includegraphics[width=1\textwidth]{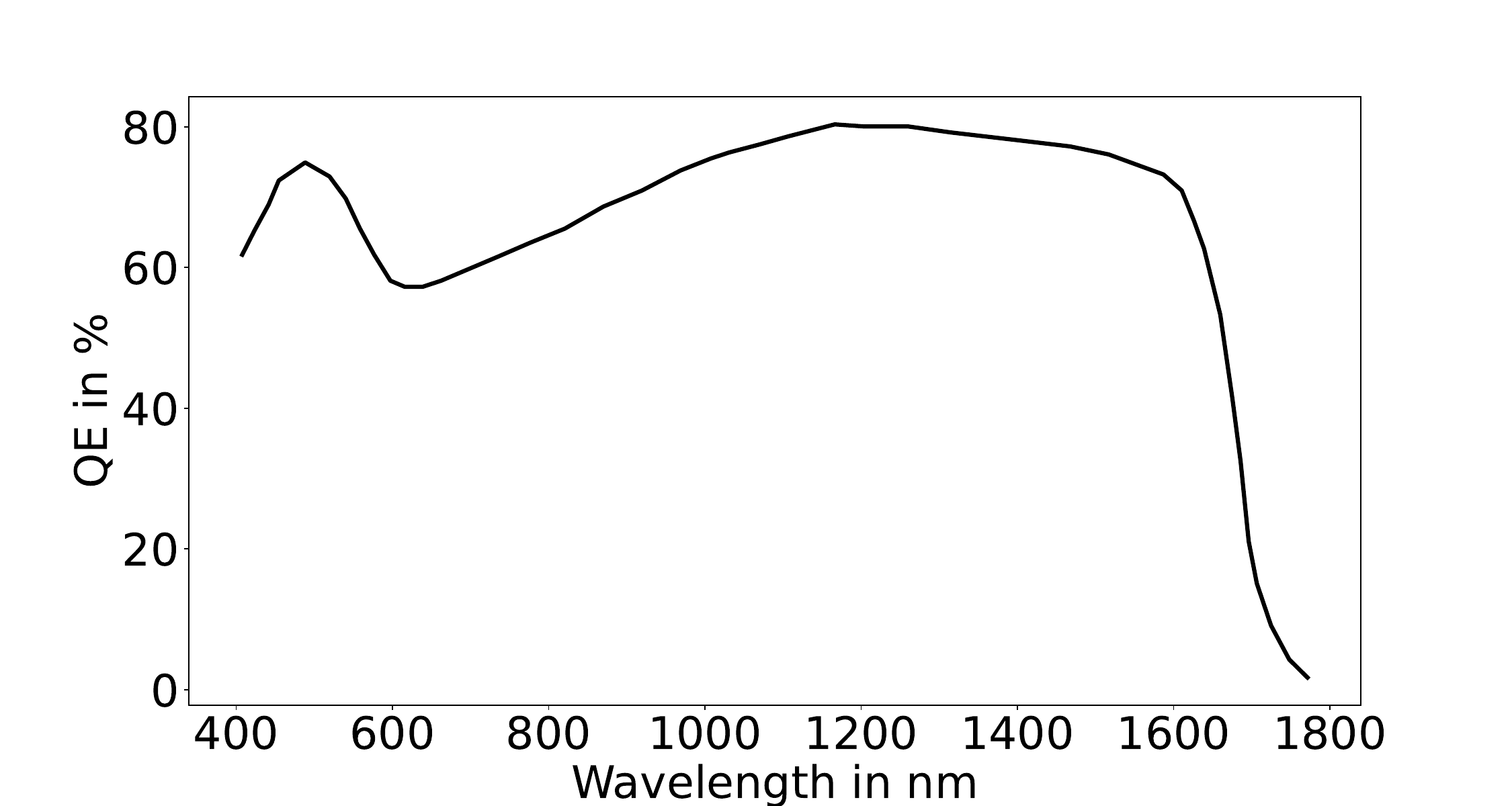}
			\caption{Quantum efficiency of IMX990 (at -20$^
   \circ $C)}\label{fig:swir_QE}
		\end{subfigure}
  
		\bigskip
  
		\begin{subfigure}{0.4\textwidth}
			
       \includegraphics[width=1\textwidth]{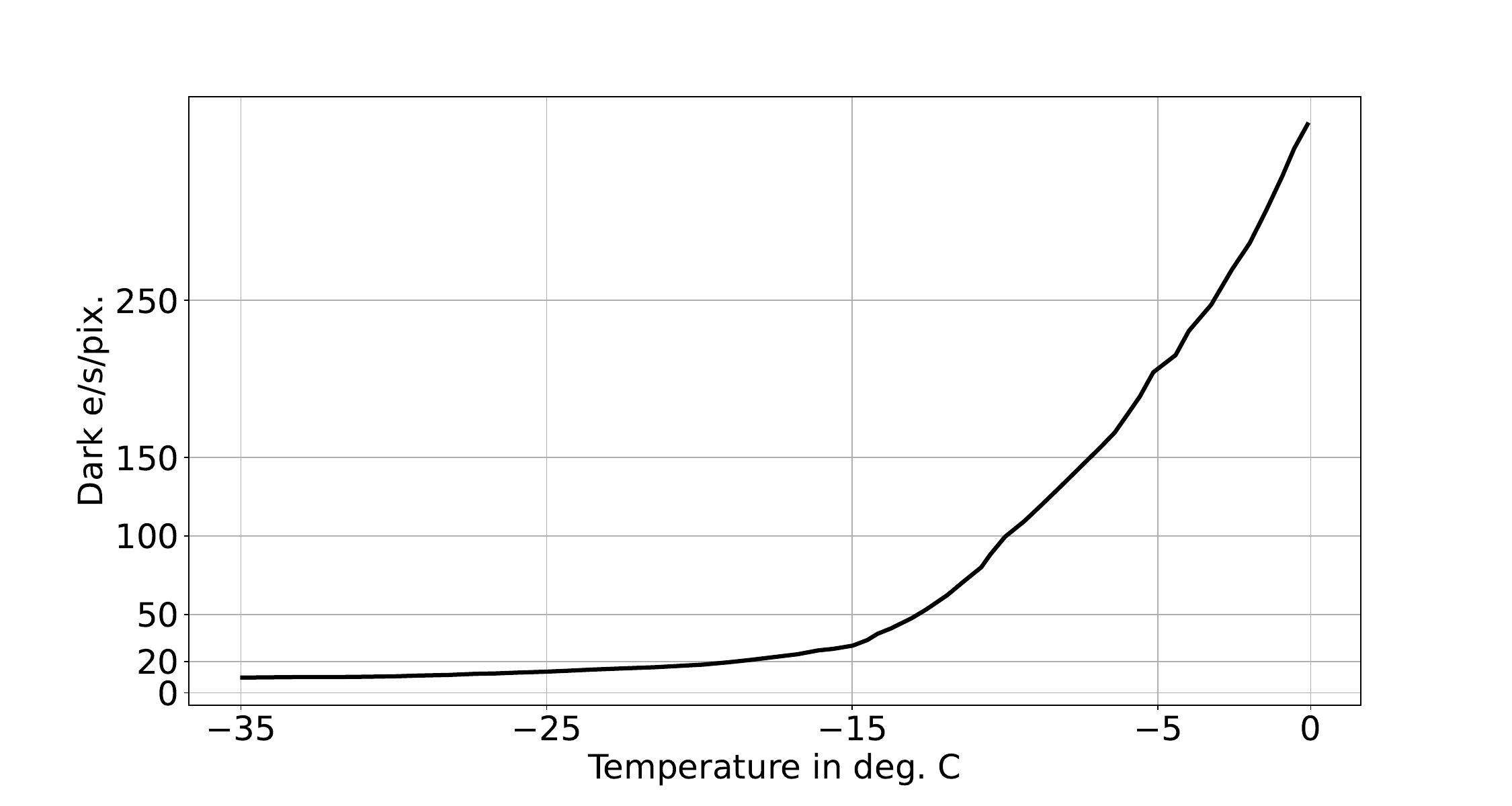}
			\caption{Dark current QHY990 camera }\label{fig:swir_dark}
		\end{subfigure}

		\caption{The quantum efficiency of the IMX990 array is shown in (a). The dark current v temperature for a QHY camera based on IMX990 is shown in (b). (Data collected from QHY website \url{https://www.qhyccd.com/} and replotted)}
		\label{fig:swir2}
\end{figure}
\begin{figure}
   \includegraphics[width=0.99\columnwidth]{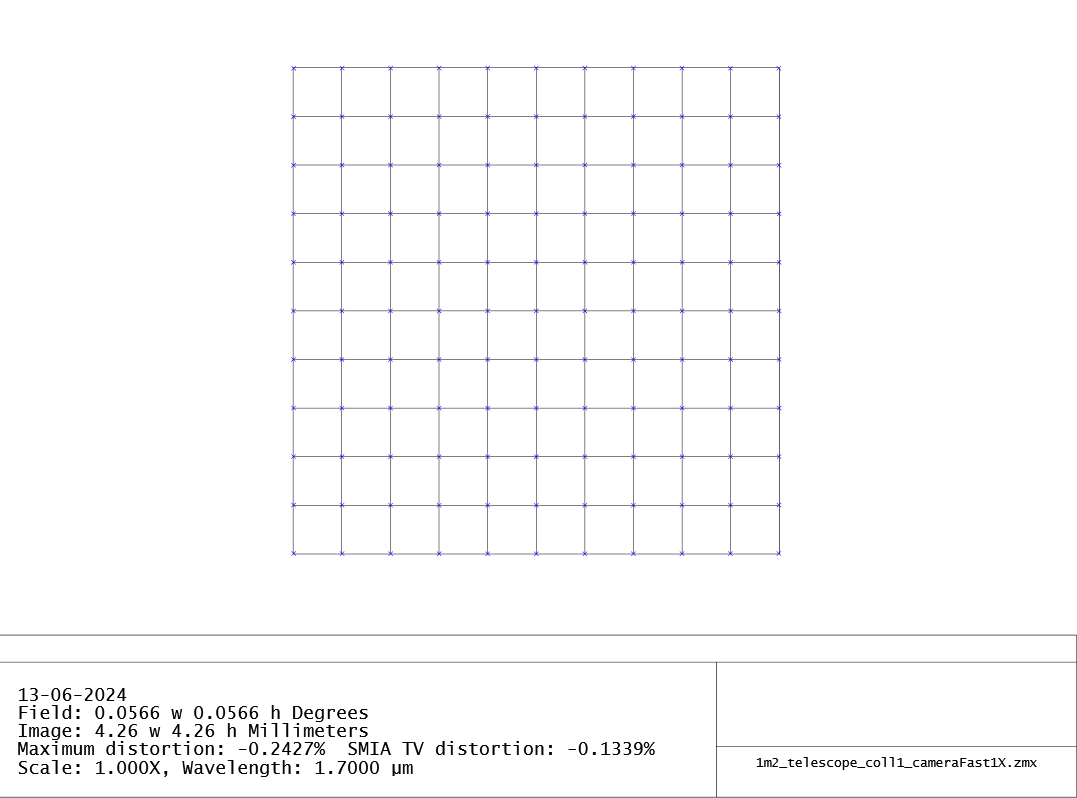}
			\caption{Grid distortion for the SWIR camera: the camera achieves fairly good distortion characteristics of about 0.25\%. This figure is obtained by referencing the centre of the achievable field of view of the instrument which is also the axis of symmetry for the instrument.}\label{fig:grid}
\end{figure}
Still, InGaAs is a developing technology, the dark noise of these arrays are typically higher compared to HgCdTe arrays based on H2RG.  Sullivan et al. (2014) \cite{sullivan2014near} have shown that for smaller telescopes, it is possible to use InGaAs arrays to achieve fairly high SNR of astronomical sources. This is particularly true for ground based instruments where photometry in SWIR wavelength range is usually background limited by strong OH backgrounds. In such a scenario, the simplicity of an InGaAs array implementation is quite lucrative. A similar analysis has been done by  \cite{mishra2022filters}  to demonstrate that it is possible to achieve high SNR ($\approx$100) within short exposure times ($<$100 seconds) for astronomical sources as faint as 12.5 magnitudes. 
 
 Additionally, present implementations of these sensors (\cite{batty2022laboratory}) show indications that these arrays do not suffer as severely from persistence issues that typically prevent HgCdTe arrays from observing brighter sources. As such, it is expected that InGaAs arrays will fill in a void seen in infrared observation of bright sources that is difficult to fulfil using only HgCdTe arrays.  The present generation of InGaAs cameras lend to simple and low cost instruments from small and medium telescopes. The lightweight  and low-complexity nature of these arrays can enable dedicated observing programs for a large number of interesting infrared sources. \\

\begin{table}[htb]
\tabularfont
\caption{SWIR camera for the 1.2M telescope}\label{tableSWIR1M2} 
\begin{tabular}{lc}
\topline

Wavelength range & 900nm to 1700 nm \\
    Field of View & 4.8 by 4 arcminutes\\
    Input F\# & F\#13 \\
    Output F\# & F\#3.6\\
    Design type & Collimator Camera\\
    Lens groups & 4 \\
    Sensor & IMX990 \\
    Barrel Length &300 mm\\
    Pixelscale & 0.25" per pixel\\
    Magnitude limit & 15 - 16 $M_J$ \\
    400 seconds integration&  \\

\hline
\end{tabular}
\tablenotes{}
\end{table}


We have selected an IMX990 (\cite{manda2019high}) based InGaAs array as the most suitable. The IMX990 sensor features a modern Cu-Cu bonding method and fairly small (5 micron square) pixel size to provide significant reduction in dark current per pixel compared to previous implementations(e.g \cite{sullivan2013precision}) of InGaAs cameras. The quantum efficiency of the array is shown in \ref{fig:swir_QE} and expected dark current levels for a QHY camera implementation is shown in \ref{fig:swir_dark}. The Sensor has usable quantum efficiencies up to 1.75 microns. This allows for observation in $Y$, $J$ and a truncated $H_S$ photometric band. The $H_S$ band has found recent use at the 2 M Liverpool telescope  (\cite{batty2022laboratory}) as well as for the Wide-field Infrared Transient Explorer(WINTER) project (\cite{lourie2020wide}) and in some cases considered an advantage as it avoids wavelength regions that are strongly affected by thermal background. \\

A collimator camera type design based on IMX990 for 1.2 M telescope of Mt. Abu observatory is presented. The 1.2 meter telescope has been a workhorse telescope for the institute since 1994 and has accommodated a number of infrared instruments over the time. The telescope has a cassegrain focus of F\#13 and has a platescale of 13.2 arc-seconds per mm. To accommodate the small pixel size of the IMX990 (5 microns) a fairly fast F\#3.6 is required at the camera output. The optical layout for the imager is shown in figure \ref{fig:layoutSony}. The design is implemented only using off-the-shelf doublets and triplets from Edmund Optics. This implementation images 4.8 $\times$ 4 arc-minutes onto the detector. The spot size performance of the design is presented in figure \ref{fig:spotingaas}. The design is expected to operate at close to the seeing limit (1"-1.5" \cite{ashok2002limited}) for the site and the telescope. The distortion of the SWIR imager is attached in figure \ref{fig:grid}.  Nominal distortion value predicted by Zemax is about 0.25\%. This figure is obtained with reference to the central field as reference. \\
  
\begin{figure}
  \centering

        \begin{subfigure}{0.49\textwidth}
         \includegraphics[width=0.99\textwidth]{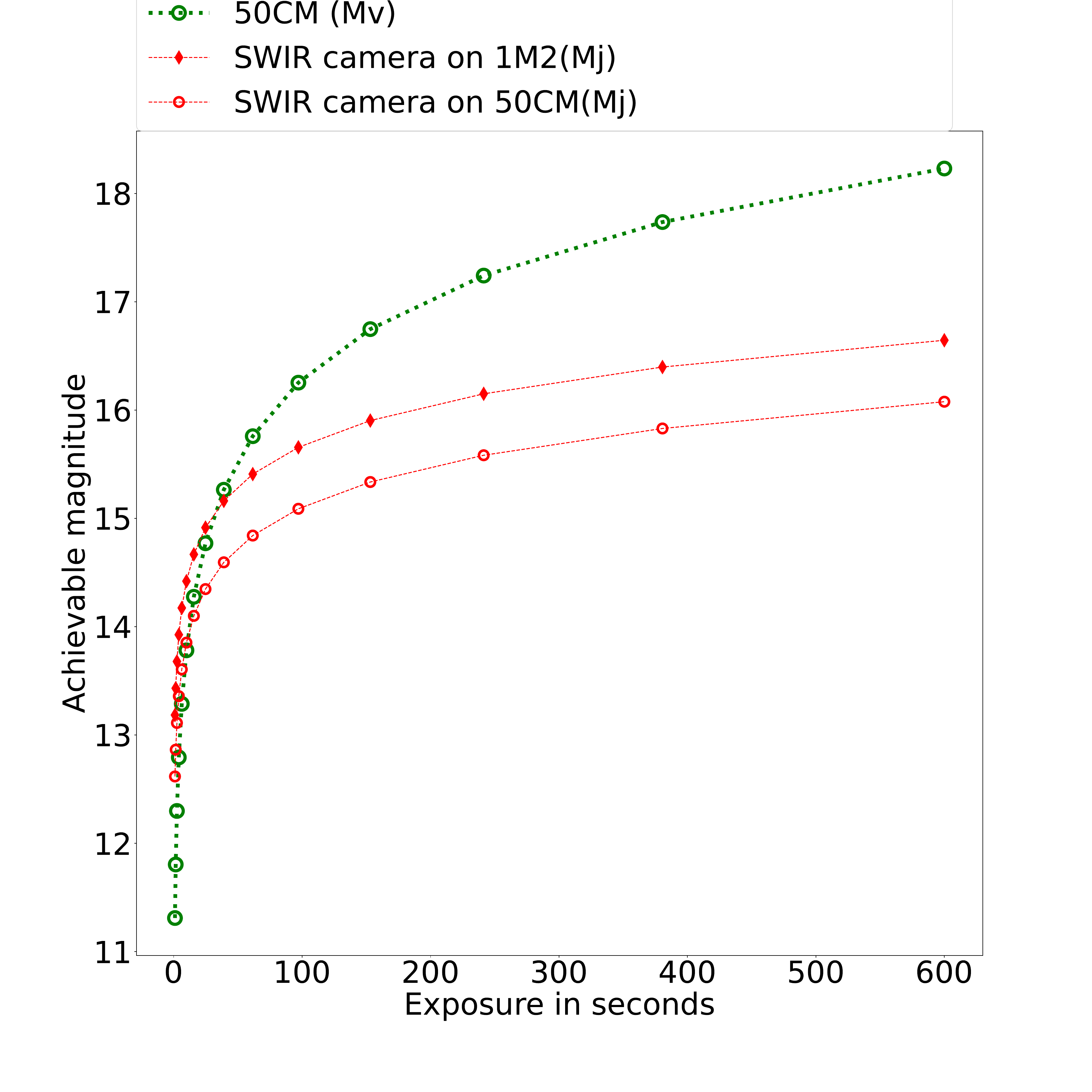}
			\caption{Expected magnitudes for different imagers}\label{fig:magimagera}
		\end{subfigure}
        \begin{subfigure}{0.45\textwidth} 
        \includegraphics[width=0.99\textwidth]{Comparison_magnitudeV.pdf}
			\caption{}\label{fig:OGLEVimager}
		\end{subfigure}
		\bigskip
		\begin{subfigure}{0.45\textwidth}
		\includegraphics[width=0.99\textwidth]{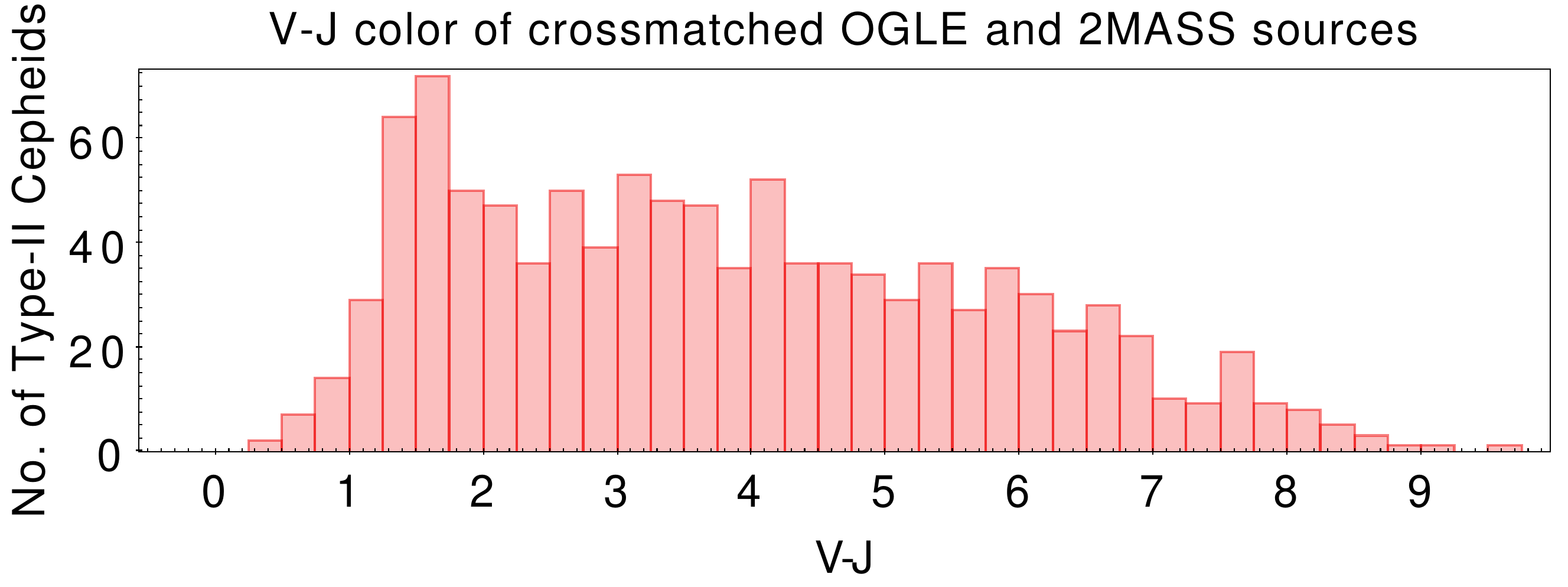}
			\caption{}\label{fig:OGLEV_J}
		\end{subfigure}

   \caption{Expected magnitudes observable through the different imagers: the expected accessible magnitudes v exposure time in seconds is plotted in (a) for different imagers discussed. The wide field camera operating in visible have a photon noise limit characteristics and the SWIR imager has a background limited characteristics. The V magnitudes of type-II cepheids in the OGLE survey is shown in (b) and their corresponding V-J colour(from a cross match with 2MASS) is shown in (c).}
   \label{fig:magimager}  
\end{figure}

The magnitude v exposure plot for this imager (along with other imagers discussed later) is shown in figure \ref{fig:magimager}. The observation at SWIR wavelengths is likely to be background limited, we have made use of the background measurements from \cite{prajapati2023near} to estimate required exposure times. As evident from figure \ref{fig:magimagera}, the accessible magnitude limit increases fairly slowly with exposure after the background level is reached (about 15 magnitude in J for Abu). Even so, this imager should be able to conduct follow up observation on a significant number of type-II cepheids and other late type variables. A histogram distribution OGLE V band magnitudes for type-II cepheids is shown in \ref{fig:OGLEVimager} and a distribution of V-J colour (by crossmatching OGLE sources with 2MASS) is shown in \ref{fig:OGLEV_J}. The distribution shows that most type-II cepheids are significantly brighter in J band than in V, this is expected as cepheids are late type stars with lower surface temperatures. We expect that the proposed SWIR imager operating on the 1.2 M telescope will be able to follow up all of GCVS catalogue sources and a significant number of OGLE sources. The imager will also serve as a supplementary facility for the upcoming Near Infrared Imager Spectrometer and Polarimter (NISP) instrument (\cite{rai2020optical}) for the 2.5 Meter telescope. By allowing the SWIR camera on the 1.2 M carry out followup and monitoring observation on brighter objects, the NISP instrument will be able to conduct focused observing runs that make use of its unique features such as  NIR polarimetry and medium resolution spectroscopy.

\subsection{Imaging cameras for the 50 CM telescope}

The 50 cm telescope at Mt. Abu is a CDK20 model from Planewave Instruments\footnote{https://planewave.com/}. The telescope has an F\# 6.8 cassegrain focus. This telescope has been extensively used to conduct dedicated observations on solar system objects (\cite{ganesh2019solar}) as well as asteroseismology (\cite{adassuriya2021asteroseismology}). Recently it has been relocated to a separate spot on the Gurushikhar peak to allow for construction of the 2.5 M telescope. Efforts are ongoing to bring this telescope back into operation. We propose for two simple cameras to utilise the facilities of the 50 CM telescope in observing variable stars. A IMX461 based ``large" format visible camera highlights the inherent capability of the telescope for wide field imaging. An SWIR camera based on the IMX990 sensor aims to provide more accessible option towards dedicated observing programs for bright infrared sources. Both of these cameras are planned to be used directly in the cassegrain focus without any need for re-imaging optics.

\begin{table}[htb]
\tabularfont
 \caption{Specifications of the two new cameras planned for the 50 CM telescope: }
\label{tab:50cm} 
 \begin{tabular}{lcc}

				\topline
				\textbf{} & \textbf{Wide field camera} & \textbf{SWIR camera} \\
				\hline
                Sensor &  IMX461 &  IMX990 \\
                
                Technology & SCMOS & InGaAs array\\
                
                Wavelength  & 400-900 nm & 500-1700 nm \\
                
                Array size & 44$\times$ 33 mm & 6$\times$5 mm\\
                
                Pixel Size &  3.76 microns   &  5 microns    \\
                FOV  & 44' $\times$ 33' & 6' $\times$ 5'  \\
                
                Achievable mag. &  $\sim17 M_V$   & $\sim15 M_J$ \\
                (200 S exposure)&&\\
                \hline
            \end{tabular}
           
		\end{table}

\subsubsection{Wide Field Visible camera\\}

The 50CM telescope has an unvignetted field of view of about 52 arc-minutes with a plate-scale of 60 arc-seconds per millimetre. This plate-scale is a good match for modern CMOS arrays which tend to have fairly small pixels($<$ 5 microns). We have selected IMX461, which is a large format camera typically aimed at the amateur astronomers yet has reasonable low noise and high QE. On the 50 CM telescope. On the 50CM telescope it images a field of view of about 44'$\times$33'. The camera implementation from QHY (model QHY461) is modest in overall weight and complexity and can be conveniently integrated into the existing mounting structure of the 50 CM telescope. The specifications for the camera are mentioned in table \ref{tab:50cm} and the expected achievable magnitudes is presented in figure \ref{fig:magimagera}.

\subsubsection{ SWIR camera\\}

 The IMX990 camera (discussed previously) has 5 micron pixels in a 1280$\times$1024 array and can also be used in the cassegrain focus of the 50 CM telescope. In this mode the array will cover a field of view of about  6'$\times$5'. Typical implementations of this sensor are fairly small and lightweight within 5 Kgs so it can be conveniently mounted on small telescopes. The expected limiting magnitude for IMX990 on the 50CM is also plotted in figure \ref{fig:magimagera}. While the limiting magnitudes achieved on the 50 CM telescope is not as  good as the implementation on the 1.2 M telescope, it will serve as a more accessible option for observing bright infrared sources.

\section{Discussions on some practical considerations}

A few practical aspects that might concern the implementation of the instruments described above are discussed as follows:

\textbf{On tolerancing of off-the-shelf components:\\}
A quick look through Edmund Optics datasheet  of doublets shows a typical  allowable ROC error of about 0.1 \% and decenter tolerance of about 1 arcminute for most of their doublets. A Zemax tolerance estimate for the Spectrograph in MonteCarlo analysis shows a mean degradation to about 14.6 micron RMS  (from a nominal of 13 microns) and 80 and 90 percentiles of about 17.5 and 19.6 microns respectively. This seems fairly acceptable as it points that the spectrograph is still going to be mostly pixel-size limited except for the worst-case scenarios (90\% and above percentiles). In this worst case, the gaps between optical elements will be used as compensators (by means of shimming the optical mounts) to recover some of the degraded performance. Of course, this might become an iterative and time consuming process but we feel this is one of the better ways to make use of off-the-shelf components.\\

\textbf{On thermal drift and focus adjustments:\\}
The smaller focal lengths (less than 150 mm) of the collimator and camera optics used in both the spectrograph as well as the SWIR imager should provide some immunity against temperature drifts.  If absolutely required, it may be preferable to move the lens assemblies rather than the detector which may be too bulky for standard precision stages. An active focusing mechanism may be realised by mounting either the collimator or the camera lens assembly on a manual linear stage such as Thorlabs MS1S. These stages typically have movement resolution of a few microns. Focus can be checked by using the calibration lamps for the spectrograph and a star for the SWIR imager. This arrangement will also have some convenience during the initial alignment phase.  If it is noticed that a large number of adjustments are required over a night then the manual stage may be upgraded to a motorised one (e.g. Thorlabs PD2). Simple steps such as insulating the optical components might also help limit the temperature swings seen by them.

 Thermal drift within the instrument is likely to be a complex issue and is best solved iteratively by gauging the extent of the actual issue and then implementing practical and feasible solutions. Hence  at this point we can only list out potential solutions rather than a definitive approach.\\

\textbf{On AR coating:\\}
By virtue of the both smaller no. of lens elements used  and also by using cemented doublets and triplets we have minimised the total glass-air interfaces and should have fairly good system throughput even without AR coating. Typically, it is possible to opt for any of the standard AR coating on most of the off-the-shelf elements from Edmund Optics (VIS, NIR I and NIR II coating options). However it is true that there is no guarantee of AR coating performance beyond the exact specification range. In this case, an uncoated version might actually be better. This will need to be worked out during the procurement of the lenses. \textbf{It is of note that the SNR calculations do not assume any AR coating at all and include the reflection losses.\\} 

\textbf{On placement of shutters, stops etc.: \\}
As both  optical designs discussed have fairly fast cameras, a conventional place for shutters is just before the first element of the camera optics. However the exact implementation of the same  is not within the scope of the present discussion as mechanical, electrical or other constraints might take priority in the final design.

In case of the SWIR camera, a pupil stop (sometimes known as a Lyot-stop) might not be necessary as the instrument is intended for observing brighter sources and excludes the wavelength regions where thermal background is prominent (i.e. beyond 1.9 microns). If absolutely necessary, the location for the pupil stop will be at the exit pupil after the collimator optics. The field stop is simply defined by the detector sensitive area at the image plane.

\section{Conclusion }

At present, work is underway to procure and install the cameras for the 50cm telescope. 
Work is ongoing to define the mechanical constraints for the other instruments vis-a-vis the different target telescopes. 
The proposed instruments will also be useful for other science cases such as observing novae and supernovae, solar system studies as well as for dedicated observing programs on late type stars. These instruments will be a value addition to the facilities present at the Mt. Abu observatory.

\appendix

\section{Magnitude V exposure calculation}

The general achievable SNR is given by \\

\begin{equation*}
   SNR = \frac{P_c(T_i)}{\sqrt{P_C(T_i) + N_R^2 + N_d \times T_i+ N_b \times T_i}}
\end{equation*}

Where,\\

\textbf{$P_c(T_i)$} is source photons collected  \\

$T_i$ is the integration time\\

$\sqrt{P_c(T_i)}$ is the photon noise \\

$N_R$ is the RMS read noise in $e^- /pixel$ \\

$N_d$ is the dark current specified in $e^- /second /pixel$. \\
The associated dark noise is $\sqrt{N_d \times T_i}$.\\ 

Similarly, $N_b$ is the background count rate specified in $e^- /second /pixel$ and the associated noise is $\sqrt{N_b \times T_i}$.

$P_c(T_i)$ is the total photons collected from a star of magnitude $M_v$, this can be expressed as

 \begin{equation*}
    P_c(T_i) = \phi _z A_{eff} T_i \times 10 ^ {(-M_v/2.5)} 
\end{equation*}

Where,\\

$\phi _z$ is the zeroth magnitude flux (\cite{zombeck2006handbook}) \\

$A_{eff}$ is the effective collecting area, and combines all the efficiencies and throughput of the system  
\begin{equation*}
   A_{eff} =  \pi \frac{D^2}{4} \times R_p  R_S (1-S_o)  Q_E  O_T  F_T  S_T
\end{equation*}

where,\\
 $D$ is the primary mirror diameter,\\
 $R_p$ is the primary mirror reflectivity, \\
 $R_s$ is the secondary mirror reflectivity, \\
 $S_o$ is the secondary obscuration factor, \\
 $Q_E$ is the detector quantum efficiency, \\
 $O_T$ is the optical throughput,\\
 $F_T$ is the nominal filter transmission,\\
 $S_T$ is the slit transmission factor (for spectroscopy)\\
 
In the case of photon noise limited observation\\
\begin{equation}
SNR = \frac{ P_c(T_i)}{\sqrt{ P_c(T_i)}} = \frac{\phi _z A_{eff} T_i \times 10 ^ {(-M_v/2.5)}}{\sqrt{\phi _z A_{eff} T_i \times 10 ^ {(-M_v/2.5)}}} 
\end{equation}

In this case the achievable magnitude$M_v$ is\\
\begin{equation}
               M_v = -2.5 log_{10}  \frac{SNR^2}{\phi _z A_{eff} T_i}
\end{equation}
This equation is valid only when high SNR( $\sim$100) is achieved. \\

For the case of SWIR photometry, often the background noise is a dominant term rather than photon noise of the source itself. \cite{sullivan2014near} have established that InGaAs based detectors do operate at the background limited scenario. \\

Then the previous  equation can be written as\\
\begin{equation}
SNR = \frac{ P_c(T_i)}{\sqrt{ P_B(T_i)}} 
\end{equation}

$\sqrt{ P_B(T_i)}$ is the background noise expressed as

$$\sqrt{\phi _z A_{eff} T_i \times 10 ^ {(-M_{B}/2.5)}}$$

 where $M_B$ is the sky background (expressed in $magnitude/arcsecond^2$). We have made use of background estimates from \cite{prajapati2023near} and a nominal seeing of 1". Then,\\
 
 $$ SNR = \frac{\phi _z A_{eff} T_i \times 10 ^ {(-M_v/2.5)}}{\sqrt{\phi _z A_{eff} T_i \times 10 ^ {(-M_{B}/2.5)}}}$$

and the achievable magnitude is expressed as\\

\begin{equation}
          M_v = -1.25log_{10}  \frac{SNR^2 \times  10 ^ {(-M_{B}/2.5)} }{(\phi _z A_{eff} ) \times T_i}
\end{equation}

This equation is also valid only when high SNR ($\sim$100) is achieved.

\section*{Acknowledgements}
Work at PRL is supported by the Department of Space, Govt. of India. 

This work has made use of OGLE database (\url{https://ogle.astrouw.edu.pl/}).

This work has made use of data from the European Space Agency (ESA) mission
{\it Gaia} (\url{https://www.cosmos.esa.int/gaia}), processed by the {\it Gaia}
Data Processing and Analysis Consortium (DPAC,
\url{https://www.cosmos.esa.int/web/gaia/dpac/consortium}). Funding for the DPAC
has been provided by national institutions, in particular the institutions
participating in the {\it Gaia} Multilateral Agreement.

This publication makes use of data products from the Two Micron All Sky Survey, which is a joint project of the University of Massachusetts and the Infrared Processing and Analysis Center/California Institute of Technology, funded by the National Aeronautics and Space Administration and the National Science Foundation.

\vspace{-1em}


\bibliography{references}



\end{document}